\documentclass[a4paper,fleqn,usenatbib]{mnras}

\usepackage[T1]{fontenc}
\usepackage{ae,aecompl}

\usepackage{graphicx}	
\usepackage{amsmath}	
\usepackage{amssymb}	

\newcommand{\nodata}{\multicolumn{1}{c}{$\cdots$}}

\title[An ALMA view of star formation in G11.92-0.61]{Simultaneous low- and high-mass star formation in a massive protocluster: ALMA observations of G11.92-0.61\thanks{This paper is dedicated to the memory of Dr. Carol Klimick Cyganowski, scholar of English, American Studies, and Theatre.}}

\author[C.~J. Cyganowski et al.]
{C.~J. Cyganowski$^{1}$\thanks{E-mail: cc243@st-andrews.ac.uk},
C.~L. Brogan$^{2}$,
T.~R. Hunter$^{2}$,
R. Smith$^{3}$,
J.~M.~D. Kruijssen$^{4,5}$,
\newauthor
I.~A. Bonnell$^{1}$,
Q. Zhang$^{6}$
\\
$^{1}$Scottish Universities Physics Alliance (SUPA), School of Physics and Astronomy, University of St. Andrews, North Haugh, \\
St Andrews, Fife KY16 9SS, UK\\
$^{2}$NRAO, 520 Edgemont Rd, Charlottesville, VA 22903, USA\\
$^{3}$Jodrell Bank Centre for Astrophysics, School of Physics and Astronomy, University of Manchester, Oxford Road, Manchester M13 9PL, UK \\
$^{4}$Astronomisches Rechen-Institut, Zentrum f\"{u}r Astronomie der Universit\"{a}t Heidelberg, M\"{o}nchhofstra{\ss}e 12-14, D-69120 Heidelberg, Germany \\
$^{5}$Max-Planck Institut f\"{u}r Astronomie, K\"{o}nigstuhl 17, 69117 Heidelberg, Germany \\
$^{6}$Harvard-Smithsonian Center for Astrophysics, 60 Garden Street, Cambridge, MA 02138, USA
}

\date{Accepted 2017 Jan 06. Received 2016 Dec 23; in original form 2016 Oct 12}

\pubyear{2017}

\begin{document}
\label{firstpage}
\pagerange{\pageref{firstpage}--\pageref{lastpage}}
\maketitle

\begin{abstract}
We present 1.05 mm ALMA observations of the deeply embedded high-mass
protocluster G11.92$-$0.61, designed to search for low-mass cores
within the accretion reservoir of the massive protostars.  Our ALMA
mosaic, which covers an extent of $\sim$0.7 pc at sub-arcsecond
($\sim$1400 au) resolution, reveals a rich population of 16 new
millimetre continuum sources surrounding the three previously-known
millimetre cores.  Most of the new sources are located in the outer
reaches of the accretion reservoir: the median projected separation
from the central, massive (proto)star MM1 is $\sim$0.17 pc.  The
derived physical properties of the new millimetre continuum sources
are consistent with those of low-mass prestellar and protostellar
cores in nearby star-forming regions: the median mass, radius, and
density of the new sources are 1.3 M$_{\odot}$, 1600 au, and n$_{\rm
  H_2}\sim$10$^{7}$ cm$^{-3}$.  At least three of the low-mass cores
in G11.92$-$0.61 drive molecular outflows, traced by high-velocity
$^{12}$CO(3-2) (observed with the SMA) and/or by H$_2$CO and CH$_3$OH
emission (observed with ALMA).  This finding, combined with the known
outflow/accretion activity of MM1, indicates that high- and low-mass
stars are forming (accreting) simultaneously within this protocluster.
Our ALMA results are consistent with the predictions of
competitive-accretion-type models in which high-mass stars form along
with their surrounding clusters.

\end{abstract}




\section{Introduction}
\label{sec:intro}

Most stars form in clusters and associations 
\citep[e.g.][]{Lada2003,Kruijssen2012}, yet even basic aspects of
how this occurs remain unknown---including the relative birth 
order of high and low mass stars.  The classic view holds that 
low-mass stars must form first because high-mass stars 
(M$>$8 M$_{\odot}$), once formed, would dissipate the natal 
cloud, preventing further star formation \citep[e.g.][]{Herbig1962}.
Some recent near-infrared, mid-infrared and X-ray studies support this view,
suggesting that distributed populations of low-mass stars may form before
high-mass stars in filamentary clouds \citep[e.g.][]{Kumar2006,Povich2010,Foster2014, Rivilla2014}.  
From optical, infrared, and X-ray observations of the DR21 massive star-forming region,
\citet{Rivilla2014} also suggest that the gravitational potential of the low-mass (proto)stellar 
population may channel mass through filaments, facilitating the formation of massive stars within the 
deepest local potential wells.  

It is, however, difficult to distinguish low-mass stars forming first from coeval low- and high-mass star formation based on observations that detect distributed populations of already-formed low-mass proto- or pre-main-sequence (PMS) stars in massive star-forming regions \citep[see e.g.][]{Foster2014}.  Studying the earliest, most embedded stages of cluster formation requires interferometric observations at (sub)millimetre wavelengths.  In the pre-ALMA era, such observations often revealed "proto-Trapezia" of closely spaced ($\lesssim$15,000 au) massive gas and dust cores,\footnote{"Core" refers to $\lesssim$0.1 pc structures likely to form single stars or small multiple systems, in contrast to "clump," which refers to parsec-scale structures of mass sufficient to form clusters.} in the absence of low-mass cores or infrared clusters \citep[e.g.\ S255N, G11.92$-$0.61, G35.03+0.35, G11.11$-$0.12;][]{Cyganowski2007,Zinchenko2012,C11sma,Brogan2011,C11vla, Wang2014}.  The ability of such studies to detect low-mass cores was, however, limited by sensitivity and dynamic range, with typical 5$\sigma$ mass sensitivities to cold (10-20 K) cores of a few solar masses.  
The two (sub)millimetre studies to date that have attained markedly better mass sensitivities (5$\sigma<<$1 M$_{\odot}$ for T=15 K) report strikingly different results.  In NGC6334I(N)--a massive (sub)millimetre (proto)cluster surrounded by infrared young stellar objects and X-ray sources--\citet{Hunter2014} detect numerous low-mass sources in 1.3 mm continuum emission, most with properties (mass and size) more suggestive of discs around low-to-intermediate mass stars than of pre- or proto-stellar cores.  In contrast, \citet{Zhang2015} do not detect a distributed population of low-mass sources in ALMA Cycle 0 observations of the infrared dark cloud (IRDC) clump and (sub)millimetre (proto)cluster G28.34$+$0.06 P1, and conclude that the low-mass stars will form later (after high-mass cluster members).  Unlike in NGC6334I(N), however, the "central" sources in P1 are not (yet) massive themselves; G28.34$+$0.06 P1 is in an early evolutionary phase, with low- to intermediate-mass embedded (proto)stars \citep[see also discussion in][]{Zhang2015}.

Establishing whether low and high mass stars \emph{form} simultaneously requires studying 
deeply embedded young proto-clusters, in which accretion is ongoing
onto high-mass (proto)stars.  Observations of this stage of protocluster evolution are
crucial for constraining models of high-mass star formation.  Theoretical models of high-mass star formation fall into two main classes: ``core accretion'' and ``competitive accretion''-type models \citep[e.g. as reviewed by][]{Tan2014}.  Core accretion models describe the collapse of self-gravitating, centrally concentrated cores with radii of $\sim$0.1 pc to form a star or small-N multiple system \citep[e.g.][and references therein]{Tan2014}.  The cores are treated as initial conditions \citep[e.g.][]{Tan2014,Myers2013,MT2003,MT2002}; as a result, this class of models makes no prediction for the formation order of high and low-mass stars within a cluster.
In contrast, competitive accretion-type models \emph{require} that high-mass stars form
in a cluster environment \citep[e.g.][]{Bonnell2004,BonnellSmith2011}.  In cluster-scale models, including those that
incorporate (proto)stellar feedback \citep[e.g.][]{Smith2009,Peters2010a,Peters2010b,Wang2010,Peters2011}, 
high-mass stars and their surrounding clusters of low and intermediate mass stars form simultaneously. 
\citet{Smith2009} show that the central, massive (proto)star accretes gas over a radius of $\sim$0.2 pc in 0.25 of the clump's dynamical time (0.25t$_{dyn}\sim$1.2$\times$10$^5$ years): within this volume there are also numerous bound, low-mass cores \citep[e.g. Figure 8 of][]{Smith2009}.  
Thus a key, testable prediction of these
models is that centrally condensed, low-mass cores should exist within the same accretion reservoir as
a forming high-mass star.

As a step towards identifying a population of low-mass cores in the vicinity of a massive millimetre (proto)cluster, we proposed an ALMA Cycle 2 observation designed to detect and characterise these objects in G11.92$-$0.61.  We chose this target because it is young, deeply embedded, and known to contain at least one accreting high-mass (proto)star.  Identified as a GLIMPSE Extended Green Object (EGO) by \citet{C08}, the extended 4.5 $\mu$m emission associated with G11.92$-$0.61 indicates the presence of shocked H$_{2}$ in an \emph{active} outflow, and hence ongoing accretion \citep[][see also \citealt{Lee2012,Lee2013}]{C08}.  Located in an IRDC \citep{C08}, the EGO is coincident with clump-scale (sub)millimetre emission in single-dish maps \citep[e.g. by][targeting \emph{IRAS} 18110-1854 $\sim$1\arcmin\/ to the NE, and in the ATLASGAL survey, \citealp{Schuller2009}]{Walsh2003,Faundez2004,Thompson2006}.
Our initial Submillimeter Array (SMA) and Combined Array for Research in Millimeter-wave Astronomy (CARMA) observations \citep[][resolution $\sim$2\farcs4 and 1\farcs1 at 1.3 mm and 1.4 mm, respectively]{C11sma} revealed three compact millimetre continuum cores, detected in thermal dust emission.
 
The brightest of these millimetre cores, MM1, is associated with a 6.7 GHz Class II CH$_{3}$OH maser \citep{C09}, strong H$_{2}$O masers \citep{HC96, Breen2011, Sato2014, Moscadelli2016}, and hot-core line emission \citep{C11sma,Cyganowski2014}, and drives a massive, high-velocity, well-collimated bipolar molecular outflow \citep[traced by $^{12}$CO(2-1), HCO$^+$(1-0), and $^{12}$CO(3-2);][]{C11sma,Cyganowski2014}.  Numerous 44 GHz Class I CH$_{3}$OH masers also trace shocked outflow gas \citep{C09}.  In 0\farcs5-resolution SMA 1.3 mm observations, a variety of hot-core molecules display consistent velocity gradients that are well-fit by a model of a Keplerian disc, including infall, that surrounds an enclosed mass of $\sim$30-60 M$_{\odot}$ \citep[of which 2-3 M$_{\odot}$ is attributed to the disc;][]{Ilee2016}.  In sum, all the known observed properties of MM1--including its weak ($\lesssim$ 1 mJy) centimetre-wavelength continuum emission \citep{C11vla,Cyganowski2014,Moscadelli2016,Ilee2016}--are consistent with this source being a massive proto-O star that is still in the process of formation, i.e.\ undergoing ongoing accretion \citep[see also][]{Ilee2016}.
 
 \begin{table*}
	\begin{minipage}{0.9\textwidth}
	\centering
	\caption{Observational Parameters.}
	\label{tab:obs}
	\begin{tabular}{lcccc} 
	\hline
    Parameter & ALMA 1.05\,mm & SMA 0.88\,mm & VLA 3\,cm & VLA 0.9\,cm \\
	\hline
	Observing date (UT) & 2015 May 14 &  2011 Aug 19 & 2015 Jun 25     & 2015 Feb 9-10 \\
	Project code        &  2013.1.00812.S &  2011A-S076  & 15A-232 & 15A-232\\ 
	Configuration       & C34-3/(4) & Extended &    A &         B \\
	Phase Centre (J2000): & & & \\
	~~~~~R.A. & 18$^{\rm h}$13$^{\rm m}$58$\fs$110$^{a}$ & 18$^{\rm h}$13$^{\rm m}$58$\fs$10 & 18$^{\rm h}$13$^{\rm m}$58$\fs$10 &  18$^{\rm h}$13$^{\rm m}$58$\fs$10 \\
	~~~~~Dec.                     & $-$18$\degr$54\arcmin22\farcs141$^{a}$ &  $-$18$\degr$54\arcmin16\farcs7 & $-$18$\degr$54\arcmin16\farcs7 & $-$18$\degr$54\arcmin16\farcs7 \\
	Primary beam size (FWHP) & n/a (mosaic) &  34\arcsec &  4\arcmin\/      & 1.3\arcmin\/ \\
	Frequency coverage: & & & & \\
	Lower band (LSB) centre & 278.2 GHz$^b$ & 335.6 GHz &  9 GHz            &  31 GHz\\  
	Upper band (USB) centre & 292.0 GHz$^b$ & 347.6 GHz &  11 GHz           &  35 GHz \\
	Bandwidth         & 2 $\times$ 1.875 GHz$^b$ & 2 $\times$ 4 GHz & 2 $\times$ 2.048 GHz & 4 $\times$ 2.048 GHz \\
	Channel spacing   & 0.9766 MHz$^b$ &   0.8125 MHz  &  1 MHz & 1 MHz \\
	                  &            &               & (30\,km\,s$^{-1}$) & (9\,km\,s$^{-1}$) \\
	Gain calibrator(s) & J1733--130 & J1733--130, J1924--292& J1832--2039 & J1832--2039 \\
	Bandpass calibrator & J1733--130 & 3C84 & J1924--2914 & J1924--2914 \\
	Flux calibrator & Titan$^{c}$   & Callisto$^{c}$ & J1331+3030 & J1331+3030 \\
	Synthesised beam$^{d}$ & 0\farcs49$\times$0\farcs34  & 0\farcs80$\times$0\farcs70  &   0\farcs30$\times$0\farcs17  & 0\farcs31$\times$0\farcs17  \\
	                         & ($\mathrm{P.A}.=-84^{\circ}$)  & ($\mathrm{P.A}.=54^{\circ}$) & ($\mathrm{P.A.}=0^{\circ}$) & ($\mathrm{P.A.}=-5^{\circ}$) \\
	Continuum rms noise$^{e}$ &  0.35 mJy beam$^{-1}$ & 3 mJy beam$^{-1}$ & 6.1 $\mu$Jy beam$^{-1}$  & 7.6 $\mu$Jy beam$^{-1}$ \\
	Spectral line rms noise$^{f}$ & 4-5 mJy beam$^{-1}$ (narrow) & 130 mJy beam$^{-1}$ ($^{12}$CO) & n/a & n/a \\
	                              & 2.5 mJy beam$^{-1}$ (wide)           &                         & & \\
	\hline
	\end{tabular}
        \begin{flushleft}
            \small{$a$: For central mosaic pointing; see Section~\ref{sec:obs_alma}.}\\  
            \small{$b$: For the two wideband spws; see Section~\ref{sec:obs_alma} for details of the five narrow spws targeting specific spectral lines.  For the wideband spws, the Hanning-smoothed spectral resolution is 1.156$\times$ the channel spacing due to online channel averaging in the ALMA correlator.}\\
	        \small{$c$: Using Butler-JPL-Horizons 2012 models.}\\  
	        \small{$d$: For continuum image.}\\  
	        \small{$e$: ALMA: the rms noise varies as a function of position in the image due to dynamic range limitations (Section~\ref{sec:obs_alma}): the quoted value is representative.} \\
	        \small{$f$: ALMA: the rms noise varies as a function of position in the image and spectral channel due to dynamic range limitations (Section~\ref{sec:obs_alma}): quoted values are for channels \emph{without} bright/complex emission, for narrow spws (imaged with $\Delta$v=0.5 km s$^{-1}$) and wide spws (imaged with $\Delta$v=1.1 km s$^{-1}$) respectively.  SMA: $^{12}$CO (3-2) data were smoothed to 1.5 km s$^{-1}$, see Section~\ref{sec:obs_other}. }
        \end{flushleft}
	\end{minipage}
\end{table*}

In addition to MM1, the massive members of the G11.92$-$0.61 (proto)cluster include MM2, a strong (sub)millimetre continuum source that lacks any other star-formation indicators and is a strong candidate for a massive starless core \citep[M$_{\rm gas}\gtrsim$30 M$_{\odot}$;][]{Cyganowski2014}.
The nature of MM3, the third known millimetre continuum core, is less clear from the SMA and CARMA observations.  MM3 is associated with a 6.7 GHz Class II CH$_{3}$OH maser and with \emph{Spitzer} MIPS 24 $\mu$m emission, both indicative of the presence of a massive (proto)star \citep{C09,C11sma}.  However, the core gas mass derived from our SMA and CARMA observations is modest ($\sim$2-9 M$_{\odot}$, depending on the assumed dust temperature), and C$^{18}$O(2-1) was the only molecule detected with the SMA \citep{C11sma}.  MM1, MM2, and MM3 are only resolved at (sub)millimetre and longer wavelengths; the total luminosity of the G11.92$-$0.61 region is $\sim$10$^{4}$ L$_{\odot}$ \citep{C11sma,Moscadelli2016}.

In this paper, we present the discovery of a population of low-mass cores in the massive (proto)cluster G11.92$-$0.61 as revealed by our sub-arcsecond-resolution ALMA Cycle 2 1.05 mm observations.  We complement our new ALMA data with deep NRAO Karl G. Jansky Very Large Array (VLA) images of G11.92$-$0.61 at 3 cm and 0.9 cm (used by \citealt{Ilee2016} and \citealt{Hunter2016} to study MM1 and MM2, respectively), and with SMA $^{12}$CO(3-2) observations that trace high-velocity outflows.  We describe our observations in Section~\ref{sec:obs}, and our results in Section~\ref{sec:results}.  Section~\ref{sec:dis} presents our analysis of the physical properties of the newly-detected sources and their molecular outflows, and discusses our results; our main conclusions are summarised in Section~\ref{sec:conclusions}.  Throughout, we adopt the maser parallax distance to G11.92--0.61 of 3.37$^{+0.39}_{-0.32}$\,kpc \citep{Sato2014}.

\begin{figure*}
	\includegraphics[width=\textwidth]{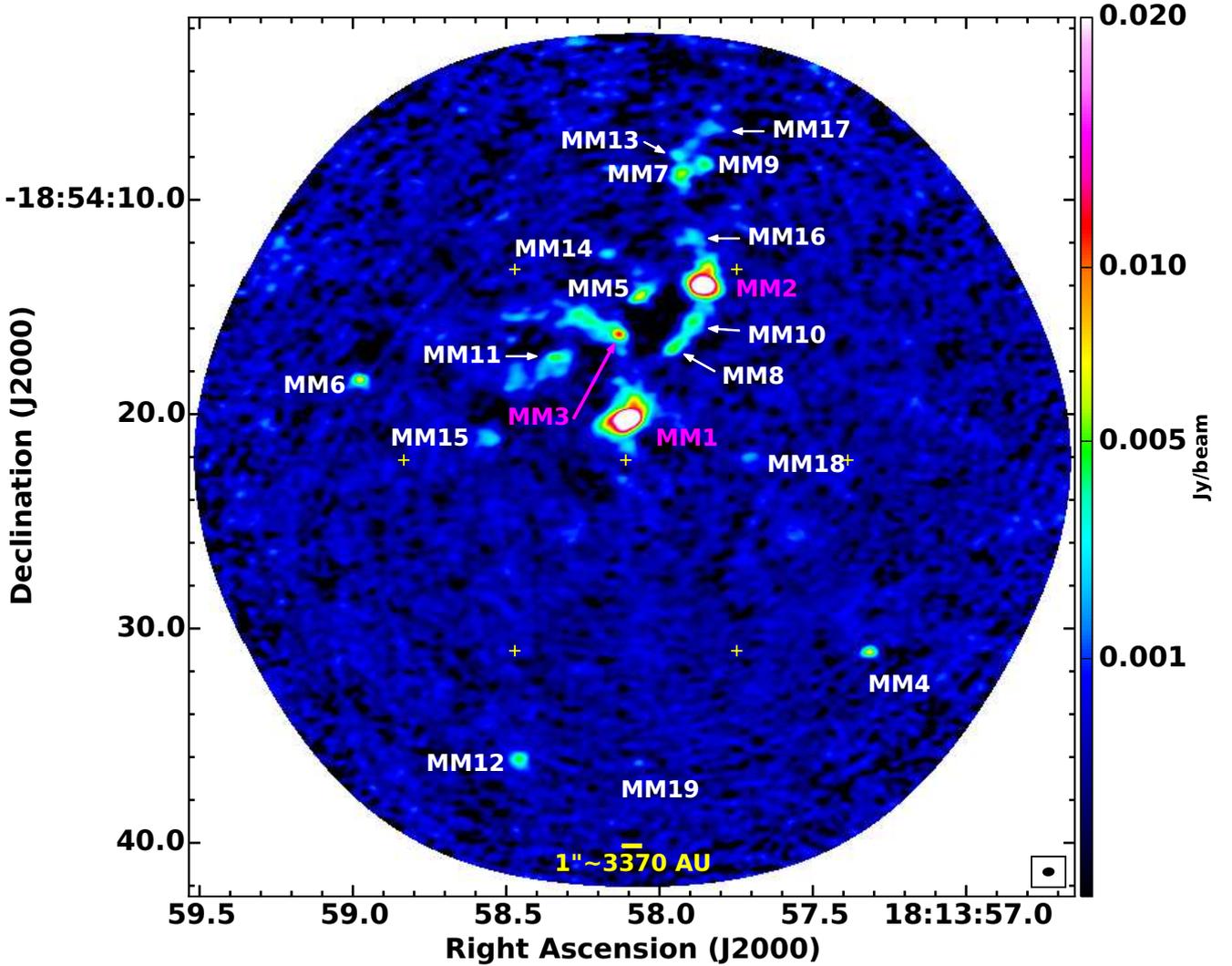}
    \caption{ALMA 1.05 mm continuum image of G11.92$-$0.61, shown in colourscale.  The ALMA synthesised beam is shown at lower right.  The edge of the colourscale image corresponds to the 30\% power point of the ALMA mosaic (width$\sim$0.65 pc, Section~\ref{sec:obs_alma}); yellow crosses mark the centres of the seven mosaic pointings.  The three previously known millimetre sources \citep[e.g.][]{C11sma} are labelled in magenta, while newly detected sources are labelled in white.  }
    \label{fig:cont_fig}
\end{figure*}

\section{Observations}
\label{sec:obs}

\subsection{ALMA}
\label{sec:obs_alma}

We observed G11.92$-$0.61 at 1.05 mm with ALMA in Cycle 2, with 37 antennas available for 
the observations.  We used a seven-pointing mosaic to cover the extent of the (sub)millimetre 
continuum emission seen with single-dish telescopes \citep[e.g.\ the ATLASGAL 
survey;][]{Schuller2009}.  The field of view of our ALMA mosaic is $\sim$40\arcsec\/ wide (within the 30\% power point; see Figure~\ref{fig:cont_fig}), equivalent to $\sim$0.65 pc at the distance of G11.92$-$0.61 (3.37 kpc; Section~\ref{sec:intro}).
The total time on-source was 44 minutes, and the projected baselines ranged from $\sim$21-542 k$\lambda$.
The largest angular scale on which the observations are sensitive to smooth structures is $\sim$8\arcsec, equivalent to $\sim$0.13 pc \citep[scale at which 10\% of peak brightness would be recovered for a Gaussian source; multiply by 0.55 for 50\% recovery;][]{Wilner1994}.
Additional observational parameters are given in Table~\ref{tab:obs}.  

The ALMA correlator was configured to cover seven spectral windows (spws): two with wide bandwidths and relatively coarse spectral resolution for continuum sensitivity (details in Table~\ref{tab:obs}), and five narrow spws tuned to particular spectral lines.  One spw covered the N$_{2}$H$^+$ (3-2) line, with a bandwidth of 468.8 MHz and a spectral resolution (Hanning-smoothed) of 0.244 MHz ($\sim$503 km s$^{-1}$ and $\sim$0.3 km s$^{-1}$, respectively).  The other four narrow spws each had 117.2 MHz ($\sim$121 km s$^{-1}$) bandwidth and 0.244 MHz spectral resolution (also Hanning-smoothed), and covered DCN (4-3) at 289.64492 GHz, $^{34}$SO 6$_7-$5$_6$ at 290.56224 GHz, H$_2$CO 4$_{0,4}-$3$_{0,3}$ at 290.62341 GHz, and C$^{33}$S (6-5) at 291.48593 GHz.   

\begin{table*}
	\begin{minipage}{0.9\textwidth}
	\centering
	\caption{Observed Properties of Continuum Sources.}
	\label{tab:obs_cont_prop}
	\begin{tabular}{lllccc} 
	\hline
    Source &  \multicolumn{2}{c}{J2000 Coordinates$^{a}$} & Peak Intensity$^{a}$ & Integ. Flux$^{a}$ & Size$^{a}$ \\
    & $\alpha$ ($^{\rm h}~~^{\rm m}~~^{\rm s}$) & $\delta$ ($^{\circ}~~{\arcmin}~~{\arcsec}$) & (mJy beam$^{-1}$) & Density (mJy) & (\arcsec $\times$ \arcsec [P.A.$^{\circ}$]) \\
	\hline
	MM1-C1$^{b}$ &  18 13 58.10792 & $-$18 54 20.2715 & 202.1 (0.9) & 245 (2) & 0.229$\times$0.144 [123] (0.008$\times$0.008 [4]) \\
	MM1-C2$^{b}$ &  18 13 58.112 &  $-$18 54 20.32 & 20.7 (0.7) & 187 (7) & 1.49$\times$0.89 [126] (0.06$\times$0.04 [3]) \\
	MM1-C3$^{b}$ & 18 13 58.109 & $-$18 54 21.50 & 1.8 (0.4) & 2.0 (0.8) & $<$0.4 \\
	MM1-total$^{c}$ & & & & 434 (2) & \\
	MM2-C1$^{b}$ & 18 13 57.8609 & $-$18 54 14.0760 & 78.8 (0.4) & 93.1 (0.8) & 0.259$\times$0.100 [103] (0.008$\times$0.007 [2]) \\
	MM2-C2$^{b}$ & 18 13 57.8601 & $-$18 54 14.088 & 27.1 (0.4) & 122 (2) & 0.90$\times$0.64 [61] (0.02$\times$0.02 [3]) \\
	MM2-C3$^{b}$ &  18 13 57.846 & $-$18 54 13.28 & 6.8 (0.4) & 30 (2) & 0.92$\times$0.56 [178] (0.07$\times$0.06 [8]) \\
	MM2-total$^{c}$ & & & & 243 (6) & \\
	MM3-C1$^{b}$ & 18 13 58.135 & $-$18 54 16.33 & 9.5 (0.7) & 14 (2) & $<$0.43$\times<$0.18 \\
	MM3-C2$^{b}$ &  18 13 58.219 &  $-$18 54 15.86 & 3.7 (0.5) & 43 (6) & 2.4$\times$0.7 [58] (0.3$\times$0.1 [4]) \\
	MM3-C3$^{b}$ &  18 13 58.29 & $-$18 54 15.7 & 1.8 (0.6) & 8 (3) & 1.6$\times$0.3 [120] (0.7$\times$0.2 [10])\\
	MM3-total$^{c}$ & & & & 49 (2) & \\
	MM4 & 18 13 57.3191 & $-$18 54 31.085 & 7.2 (0.3) &  8.6 (0.6) & 0.22$\times$0.14 [96] (0.07$\times$0.08 [66]) \\
	MM5 & 18 13 58.062 & $-$18 54 14.50 & 7.0 (0.6) & 18.0 (1.9) & 0.78$\times$0.27 [130] (0.11$\times$0.08 [7])\\
	MM6 & 18 13 58.979 & $-$18 54 18.493 & 6.8 (0.4) & 11.1 (0.9) & 0.37$\times$0.28 [70](0.08$\times$0.07 [55]) \\
	MM7 & 18 13 57.928 & $-$18 54 08.88 & 5.6 (0.6) & 25 (3) & 0.9$\times$0.6 [151] (0.1$\times$0.1 [21]) \\
	MM8 & 18 13 57.949 &  $-$18 54 16.87 & 5.3 (0.4) & 18 (2) & 1.06$\times$0.28 [133] (0.11$\times$0.06 [3])\\
	MM9 & 18 13 57.861 & $-$18 54 08.45 & 4.5 (0.4) & 13 (2) & 0.6$\times$0.5 [132] (0.1$\times$0.1 [85]) \\
	MM10 & 18 13 57.891 & $-$18 54 15.79 & 4.5 (0.3) & 22 (2) & 1.16$\times$0.50 [144] (0.10$\times$0.06 [5])\\ 
	MM11$^{d}$ & 18 13 58.336 & $-$18 54 17.45 & 4.5 (0.3) & 12 (1) & 0.81$\times$0.33 [98] (0.10$\times$0.05 [5]) \\
	MM12 & 18 13 58.460 & $-$18 54 36.10 & 4.4 (0.3) & 13 (1) & 0.61$\times$0.48 [29] (0.09$\times$0.10 [39]) \\
	MM13 &  18 13 57.935 &  $-$18 54 08.06 & 3.1 (0.5) & 10 (2) & 0.7$\times$0.5 [30] (0.2$\times$0.2 [36]) \\
	MM14 & 18 13 58.172 & $-$18 54 12.58 & 2.8 (0.3) & 5.3 (0.8) & 0.49$\times$0.29 [87] (0.12$\times$0.09 [33]) \\
	MM15 & 18 13 58.563 & $-$18 54 21.20 & 2.6 (0.3) & 10 (2) & 0.8$\times$0.6 [99] (0.2$\times$0.1 [86]) \\
	MM16 & 18 13 57.895 & $-$18 54 11.90 & 2.4 (0.3) & 15 (2) & 1.1$\times$0.8 [64] (0.2$\times$0.2 [27]) \\
	MM17 & 18 13 57.845 & $-$18 54 06.81 & 2.2 (0.3) & 11 (2) & 1.2$\times$0.5 [107] (0.3$\times$0.1 [11]) \\
	MM18 & 18 13 57.707 & $-$18 54 22.14 & 1.7 (0.2) & 5.5 (0.9) & 0.8$\times$0.5 [85] (0.2$\times$0.1 [31]) \\
	MM19 &  18 13 58.071 & $-$18 54 36.28 & 1.6 (0.2) & 1.9 (0.5) & $<$0.4 \\
	\hline
	\end{tabular}
        \begin{flushleft}
            \small{$a$: From two-dimensional Gaussian fitting; ``size'' is deconvolved source size (FWHM).  Statistical uncertainties are indicated by the number of significant figures or given in parentheses.  For sources too small/weak to have reliable fitted sizes, "size" is quoted as $<$ the geometric mean of the synthesised beam.}\\  
            \small{$b$: Source fit with multiple Gaussian components, designated -C\#.  To obtain a reasonably good fit, MM1-C3 was fit separately (while MM1-C1 and MM1-C2 were fit simultaneously, with a two-component Gaussian fit) and a zero-level offset of 0.77 mJy beam$^{-1}$ was required.  MM3-C1 is fit as a point source: the size quoted is the upper limit from the CASA IMFIT task.  MM3-C2 and MM3-C3 together describe the extended emission NE of the MM3-C1 point source.}\\
            \small{$c$: Flux density measured within the local 3$\sigma$ contour level; quoted uncertainties are estimated as the local rms multiplied by the number of independent beams within the source area.  For MM2, the 3$\sigma$ contour level connects MM2 to MM8, MM10, and MM16.  Instead of reporting a total flux that would encompass all of these sources, we additionally apply RA and Dec limits (RA: 18$^{\rm h}$13$^{\rm m}$57.75$^{\rm s}$ to 18$^{\rm h}$13$^{\rm m}$57.94$^{\rm s}$, Dec: $-$18$^{\circ}$54\arcmin12.5\arcsec\/ to $-$18$^{\circ}$54\arcmin15.3\arcsec) to isolate MM2, and increase the quoted uncertainty to account for the uncertainty introduced by the choice of limits.}\\
            \small{$d$: Fit with a zero-level offset of 0.37 mJy beam$^{-1}$ to obtain a reasonable fit to the dominant compact source; it is unclear if the weaker emission to the south is associated with MM11, or arises from additional sources not resolved in our data.  }
        \end{flushleft}
	\end{minipage}
\end{table*}

The ALMA data were calibrated using the CASA 4.2.2 version of the ALMA calibration pipeline.  
Titan was used to set the absolute flux scale, and we estimate the absolute flux uncertainty to be
$\sim$10\%.  
After the calibration was applied, the science target data were split off, and line-free channels were used to construct a pseudo-continuum dataset.  Identifying line-free channels is complicated by the line-rich spectrum of the hot core MM1.  To select line-free channels for continuum imaging and continuum subtraction, we first made dirty line+continuum cubes of the seven ALMA spws.  Using these cubes, we tested different threshold levels for line-free channel identification (this approach is similar to that used by \citealt{Brogan2016} in their study of NGC6334I, with two line-rich hot cores).
Selecting line-free channels in the C$^{33}$S spectral window (spw 3) proved problematic due to wide lines and possible absorption features, and we omitted this narrow spw from our final continuum dataset. 
Based on our threshold tests and comparison of the flux densities in the resulting continuum images, we estimate a residual line contamination for the hot core source MM1 at the level of 10-15\% for the adopted threshold level.  The adopted threshold represents a compromise between reducing line contamination of MM1 and attaining good sensitivity to weak continuum sources with little or no line emission, since the image rms noise depends on the aggregate continuum bandwidth.  The total "line-free" bandwidth used for our final continuum image is $\sim$2.4 GHz.   

The continuum data were iteratively self-calibrated, and the solutions applied to the line data.
The final ALMA continuum image was made using multi-frequency synthesis and a robust weighting parameter of 0.  This yields a synthesised beam of 0\farcs49$\times$0\farcs34, equivalent to a linear resolution of $\sim$1650$\times$1150 au at 3.37 kpc.  The rms noise in the ALMA continuum image varies as a function of position due to dynamic range limitations associated with the bright continuum sources (in particular MM1).  The dynamic range of our ALMA image is $\sim$1000 (peak/minimum rms), but the measured rms varies by more than a factor of two across the image, from $\sim$0.25 mJy beam$^{-1}$ to $\gtrsim$0.5 mJy beam$^{-1}$ (in the immediate vicinity of MM1 and MM3).     
For transitions with bright emission associated with MM1 or its outflow, the line data are also affected by dynamic range limitations and increased noise associated with poorly-recovered extended structure(s).  For H$_2$CO 4(0,4)-3(0,3) and N$_{2}$H$^+$ (3-2), the rms in channels with bright, complex emission is 2-5$\times$ higher than the $\sim$4 mJy beam$^{-1}$ measured in channels without bright emission.   
The line image cubes were made with a robust weighting parameter of 0.5, to achieve the best compromise between sensitivity and confusion from missing short spacing information.  The synthesised beamsize varies slightly for the various line image cubes, as a function of frequency: for example, the synthesised beam is 0\farcs54$\times$0\farcs40 (P.A.=$-$82$^{\circ}$) for the N$_{2}$H$^+$ (3-2) cube ($\nu_{\rm rest}=$279.51175 GHz), and 0\farcs53$\times$0\farcs38 (P.A.=$-$83$^{\circ}$) for the H$_2$CO 4$_{0,4}-$3$_{0,3}$ cube.  For better sensitivity to faint emission, the narrow spws were imaged with 0.5 km s$^{-1}$ channels.   
All measurements were made from images corrected for the primary beam response.

\subsection{SMA and VLA}
\label{sec:obs_other}

To complement our new ALMA observations, we make use of previously-published SMA 0.88 mm and VLA 3 cm and 0.9 cm
observations of G11.92$-$0.61.  The SMA 0.88 mm dataset was used by \citet{Cyganowski2014} to study the candidate massive
prestellar core G11.92$-$0.61 MM2, and is described in that publication.  Observational details are included in
Table~\ref{tab:obs} for completeness; here, we consider only the $^{12}$CO(3-2) data.  To better identify and
characterise high-velocity outflows from low-mass (proto)stars, we have reimaged the $^{12}$CO(3-2) data to generate an image cube smoothed to 1.5 km s$^{-1}$ resolution \citep[as compared to 3 km s$^{-1}$ in][]{Cyganowski2014}.  The spectral line rms quoted in Table~\ref{tab:obs} is for the new 1.5 km s$^{-1}$ cube.  The $^{12}$CO(3-2) synthesised beam
is 0\farcs80$\times$0\farcs72 (P.A.=45$^{\circ}$), $\sim$2700$\times$2400 au at 3.37 kpc.  
The VLA 3 cm and 0.9 cm data were used by \citet{Ilee2016} to study
G11.92$-$0.61 MM1, a candidate Keplerian disc around a proto-O star, and are described in that publication.  Details of 
these VLA observations are also presented in Table~\ref{tab:obs}.

\begin{figure*}
	\includegraphics[width=\textwidth]{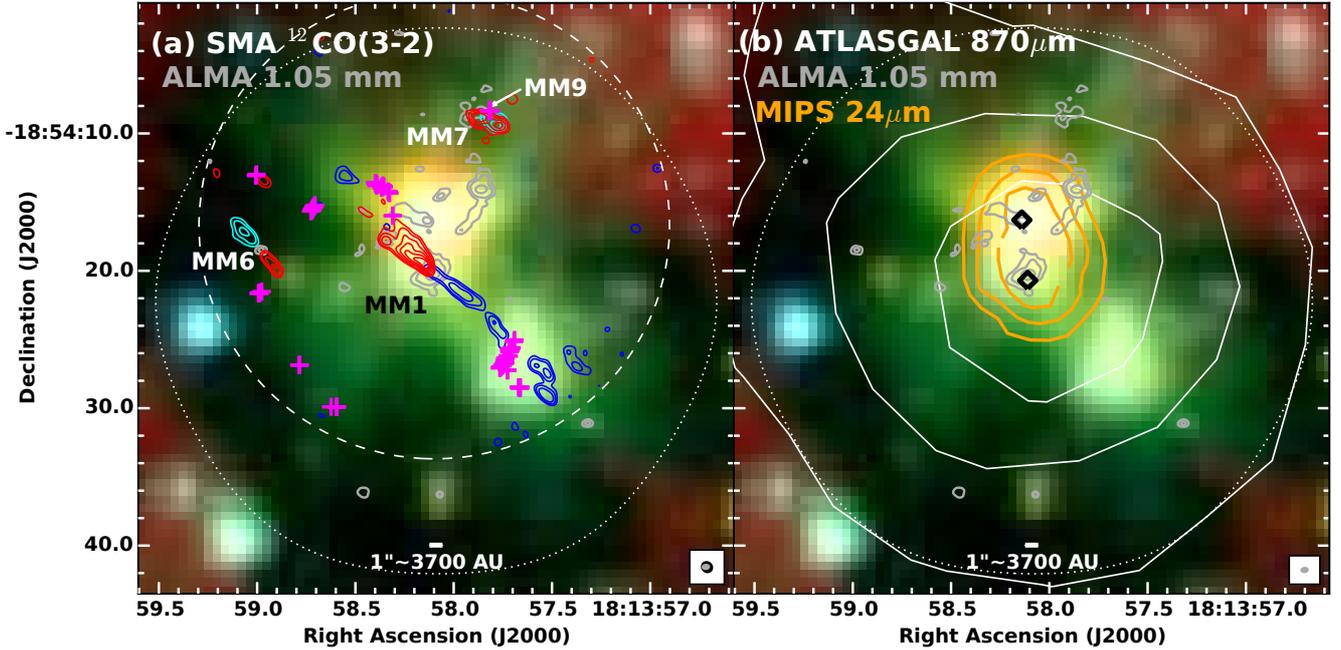}
    \caption{\emph{Spitzer} GLIMPSE three-colour image (RGB: 8.0, 4.5, 3.6 $\mu$m) overlaid with ALMA 1.05 mm continuum contours (dark grey) and contours of (a) composite maps of blueshifted/redshifted SMA $^{12}$CO(3-2) emission (corrected for the response of the primary beam, and masked outside its FWHP level); (b) ATLASGAL 870 $\mu$m (white) and \emph{Spitzer} MIPS 24 $\mu$m (orange) emission.  In both panels, the 30\% power point of the ALMA mosaic (width$\sim$0.65 pc, Section~\ref{sec:obs_alma}) is shown as a dotted white contour; in (a), the FWHP SMA primary beam is also shown, as a dashed white circle.  
    Class I CH$_{3}$OH masers (magenta $+$) are marked in (a), Class II CH$_{3}$OH masers (black $\Diamond$) in (b) \citep[both from][]{C09}.
    Contour levels: ALMA 1.05 mm: [5,15,100]$\times$0.35 mJy beam$^{-1}$--note that with these levels MM19 is not contoured (Section~\ref{sec:obs_alma}, Tables~\ref{tab:obs} and \ref{tab:obs_cont_prop}), a grey $\circ$ marks its position from Table~\ref{tab:obs_cont_prop}; SMA $^{12}$CO: 1.0 Jy beam$^{-1}$ km s$^{-1}$ $\times$ [4,6,9] (blue, cyan), $\times$ [4,6,9,12,15,18] (red).  Velocity ranges are optimised for each source (see Section~\ref{sec:line_results}). 
    ATLASGAL: [0.4, 0.6, 0.8]$\times$peak=5.25 Jy beam$^{-1}$; MIPS: 900, 1350, 1800 MJy sr$^{-1}$.
    The ATLASGAL and MIPS resolutions are $\sim$19\arcsec\/ and $\sim$6\arcsec, respectively;
     each panel shows the ALMA (grey) and/or SMA (black) synthesised beam(s) at the lower right.      }
    \label{fig:irac_CO}
\end{figure*}

\section{Results}
\label{sec:results}

\subsection{Continuum Emission}
\label{sec:cont_results}

Figure~\ref{fig:cont_fig} presents our ALMA 1.05 mm continuum image of G11.92$-$0.61.  As shown 
in Fig.~\ref{fig:cont_fig}, the sensitivity and imaging fidelity of ALMA reveal numerous 
additional sources surrounding the three massive (proto)cluster members.  Within our ALMA mosaic 
(30\% power point) we detect 16 new sources at $>$5$\sigma$, which we name MM4...MM19 in order of
descending peak intensity.  As noted in Section~\ref{sec:obs_alma}, the rms noise varies 
significantly across the image; we measured the rms locally to evaluate whether potential 
sources met the $>$5$\sigma$ threshold.
We carefully examined our deep VLA 3 cm and 0.9 cm images (Table~\ref{tab:obs}, Section~\ref{sec:obs_other}) to check for centimetre-wavelength counterparts to the newly discovered ALMA sources.  None of MM4...MM19 are detected in the VLA images, with 4$\sigma$ limits of 24.4 $\mu$Jy beam$^{-1}$ at 3 cm and 30.4 $\mu$Jy beam$^{-1}$ at 0.9 cm.

The observed properties of the millimetre continuum sources, as determined from two-dimensional Gaussian fits, are presented in Table~\ref{tab:obs_cont_prop}.  The newly detected sources are sufficiently compact that a single two-dimensional Gaussian provides a reasonable fit to the observed emission.  
In contrast, the previously known millimetre sources MM1, MM2, and MM3 require multiple Gaussian components to obtain reasonably noise-like residuals (Table~\ref{tab:obs_cont_prop}).
With our current angular resolution, it is unclear whether any of MM1..MM3 harbor multiple cores or protostars, or if these millimetre sources simply have non-axisymmetric extended dust emission.  For this reason, we do not consider components designated -C* as part of our newly discovered source population.   
We also note that for MM1, even a multiple-component fit leaves significant residuals (the strongest with a peak of $\sim$7 mJy beam$^{-1}$), which may indicate the presence of additional compact structure(s) unresolved by our observations.  
Recent ALMA observations of NGC6334I with 220 au resolution ($\sim$6$\times$ finer than the G11.92$-$0.61 data presented here) reveal that the dominant hot core NGC6334I-MM1 is comprised of seven components clustered within a radius of 1000 au \citep{Brogan2016}.
Higher angular resolution observations will be required to probe the substructure of the bright sources MM1..MM3 in G11.92$-$0.61.   
For these three sources, Table~\ref{tab:obs_cont_prop} presents a "total" flux density measured within the local 3$\sigma$ contour level, for comparison with the Gaussian fitting results.   

Figure~\ref{fig:irac_CO} shows our ALMA continuum results in the context of previous observations of G11.92$-$0.61 at mid-infrared (MIR)-centimetre wavelengths.  As seen in Figure~\ref{fig:irac_CO}, the Extended Green Object G11.92$-$0.61 has a bipolar MIR morphology \citep[see also e.g.][]{C08,C09,C11sma}: the northeast lobe is bright in multiple \emph{Spitzer} IRAC bands and at 24 $\mu$m (in addition to its 4.5 $\mu$m emission), while the southwest 4.5 $\mu$m lobe is associated with strong 44 GHz Class I CH$_{3}$OH masers, which trace shocked gas at outflow-cloud interfaces \citep[e.g.][]{Plambeck1990,Kurtz2004,C09}.
MM1 and MM3 are coincident with the MIR-bright northeast lobe, as are a few of the newly-detected millimetre continuum sources (MM5, MM11, and MM14).  The ridge or filament made up of MM8 and MM10 lies near the western edge of the northeast lobe, and is also coincident with multiband MIR emission.  These two sources are unusual among the new ALMA detections in forming part of a larger, elongated millimetre continuum structure, and also in their associated line emission (Section~\ref{sec:line_results}).  
In massive star-forming regions, H$_2$ (from shocked gas in outflows) and PAHs (easily excited by massive (proto)stars) contribute significantly to the IRAC bands \citep[see e.g. discussion in][]{C08}. We note that in this context, and given the comparatively coarse resolution of IRAC ($\sim$1.5-2\arcsec) relative to our ALMA observations, the fact that MM5, MM8, MM10, MM11, and MM14 are coincident with the northeast lobe does not necessarily mean that these sources are themselves sufficiently warm or energetic to excite MIR emission.
The (saturated) MIPS 24 $\mu$m emission is likely a blend of emission from MM1 and MM3 \citep[see also discussion in][]{C09,C11sma}, each of which is associated with a 6.7 GHz Class II CH$_3$OH maser, indicative of the presence of a massive (proto)star \citep[e.g.][]{Minier2003}.  The nature of MM3 is discussed further in Section~\ref{sec:dis_mm3}.  
 
The majority of the newly-detected millimetre continuum sources are located on the fringes of the GLIMPSE Extended Green Object, coincident either with extended 4.5 $\mu$m emission (which appears as green in Figure~\ref{fig:irac_CO}) or with no MIR emission at all.   
Three sources (MM4, MM15, and MM18) are coincident with 4.5 $\mu$m emission, while two (MM6 and MM12) are MIR-dark.
The northern group of sources comprised of MM7, MM9, MM13, and MM17 lies between two fingers of extended 4.5 $\mu$m emission, as does MM16.
Interestingly, three of the new millimetre continuum sources detected with ALMA--MM4, MM6, and MM12--are notably isolated, and are either point sources (MM4 and MM6) or approximately round (MM12).  Of these three, MM6 and MM12 are located in infrared-dark regions of the \emph{Spitzer} GLIMPSE image, e.g.\ in the IRDC. 
Notably, while many of the newly-detected millimetre sources are located far from the cluster centre (e.g. from MM1; see also Section~\ref{sec:dis_phys_prop}) and from bright MIR emission, all are within the half-power level of the ATLASGAL 870 $\mu$m clump (Figure~\ref{fig:irac_CO}b).

We note that MM19 is an outlier among the ALMA-detected sources, in both its MIR and millimetre properties.  The weakest of the ALMA detections (S$_{\rm int}$=1.9$\pm$0.5 mJy), MM19 is also the only new source too small/weak to obtain a reliable fit for its size (Table~\ref{tab:obs_cont_prop}).
Indeed, MM19 meets our 5$\sigma$ threshold only because it is located in the area of the ALMA continuum image with the lowest rms noise.
In the MIR, MM19 is coincident with an IRAC source that appears green in Figure~\ref{fig:irac_CO}, but more point-like than the majority of the 4.5 $\mu$m emission associated with the Extended Green Object.
MM19 is also not located along the NE-SW MM1 outflow axis, making it less likely that the 4.5 $\mu$m counterpart to MM19 is associated with this dominant outflow.  This combination of properties raises the possibility that MM19 may differ in character from the other newly detected millimetre continuum sources; the nature of the newly-detected sources is discussed further in Section~\ref{sec:dis_phys_prop}.

\begin{figure*}
	\includegraphics[width=\textwidth]{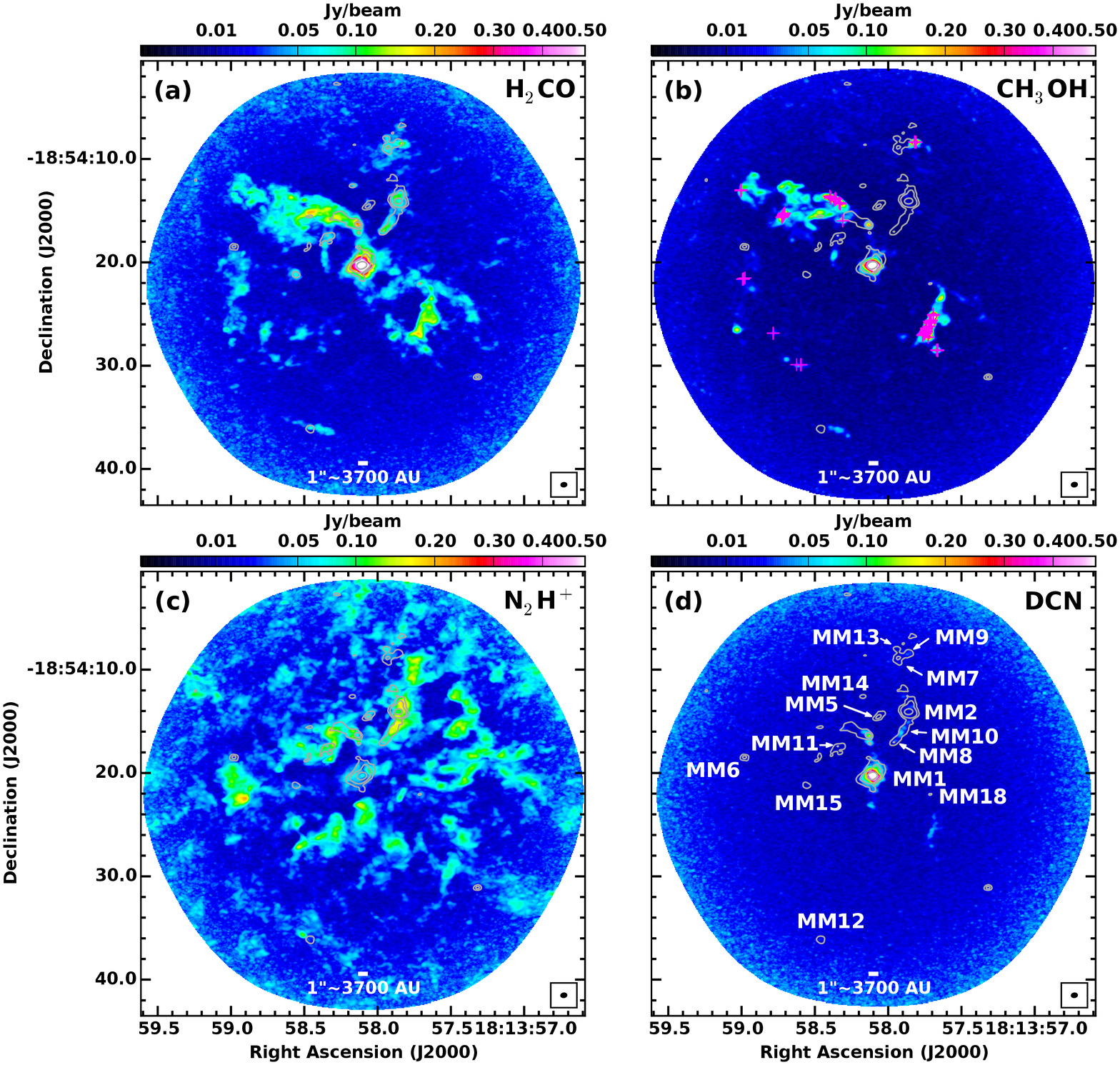}
    \caption{ALMA peak maps of (a) H$_2$CO 4$_{0,4}-$3$_{0,3}$, (b) CH$_{3}$OH $9_{-1}-8_{0}$ (Class I maser line; \citealt{Voronkov2012,ALMA_278_maser}), (c) N$_{2}$H$^+$ (3-2), and (d) DCN (4-3) in colourscale, overlaid with contours of ALMA 1.05 mm continuum emission (levels: [5,15,100]$\times$0.35 mJy beam$^{-1}$; note that with these levels MM19 is not contoured, see Section~\ref{sec:obs_alma} and Tables~\ref{tab:obs} and \ref{tab:obs_cont_prop}).  In each panel, the ALMA synthesised beam is shown at lower right, and the edge of the colourscale image corresponds to the 30\% power point of the ALMA mosaic.  In (b), positions of 44 GHz Class I CH$_{3}$OH masers from \citet{C09} are marked with magenta $+$.    }
    \label{fig:lines_4panel}
\end{figure*}

\subsection{Line Emission}
\label{sec:line_results}

In this paper, we focus on molecular line emission as it relates to the newly-discovered population of millimetre continuum sources; the copious, compact hot-core line emission associated with MM1 will be the subject of a future publication, and is not considered here.     
Our initial millimetre-wavelength interferometric observations of G11.92$-$0.61 showed a single dominant bipolar molecular outflow, driven by MM1 and traced by high-velocity, well-collimated $^{12}$CO(2-1) and HCO$^+$(1-0) emission \citep[in 2\farcs4-resolution 1.3 mm SMA and 5\farcs9-resolution 3 mm CARMA observations, respectively;][]{C11sma}.  
Figure~\ref{fig:irac_CO}a presents this outflow--centred on MM1, with the redshifted lobe to the NE and the blueshifted lobe to the SW--as seen at 0\farcs76 ($\sim$2550 au) resolution with the SMA in $^{12}$CO(3-2)  \citep[Section~\ref{sec:obs_other}, Table~\ref{tab:obs}; see also][]{Cyganowski2014,Ilee2016}.
Remarkably, combining the subarcsecond-resolution SMA $^{12}$CO(3-2) data with our new ALMA 1.05 mm continuum image reveals that at least two of the new ALMA sources--MM6, and MM7 and/or MM9--are also driving molecular outflows.  

As shown in Figure~\ref{fig:irac_CO}a, the outflow driven by MM6 is extremely well-collimated; the blue- and red-shifted lobes are well separated, and centred on the millimetre continuum source.  Blue- and red-shifted $^{12}$CO(3-2) emission is also detected in the vicinity of MM7 and MM9, coincident with 44 GHz Class I CH$_3$OH masers.  In this case, there is substantial overlap between the blue- and red-shifted lobes; it is unclear whether MM7 or MM9 is the more likely driving source, and indeed whether the CO emission is attributable to one or to multiple outflows.  
The red, blue, and cyan contours in Figure~\ref{fig:irac_CO}a represent a composite map of the blue- and redshifted $^{12}$CO(3-2) emission: integrated velocity ranges are optimised for each outflow, assuming v$_{\rm LSR}$=35 km s$^{-1}$ (in the absence of compact emission from, and so measured velocities for, the newly-detected millimetre continuum sources, as discussed below).  The integrated velocity ranges for the MM6 outflow are 6.5--27.5 km s$^{-1}$ (blue) and 48.5--84.5 km s$^{-1}$ (red), for the MM7/MM9 outflow are 23.0--30.5 km s$^{-1}$ (blue) and 42.5--56.0 km s$^{-1}$ (red), and for the remainder of the map (including the main MM1 outflow) are $-$17.5--20.0 km s$^{-1}$ (blue) and 50.0--74.0 km s$^{-1}$ (red).

Figure~\ref{fig:lines_4panel} presents peak maps of the ALMA H$_2$CO 4$_{0,4}-$3$_{0,3}$ ($\nu_{\rm rest}$=290.62341~GHz, E$_{\rm upper}$=35 K), CH$_{3}$OH $9_{-1}-8_{0}$ ($\nu_{\rm rest}$=278.30451 GHz, E$_{\rm upper}$=110 K), N$_{2}$H$^+$ (3-2) ($\nu_{\rm rest}$=279.51175 GHz, E$_{\rm upper}$=27 K), and DCN (4-3) ($\nu_{\rm rest}$=289.64492 GHz, E$_{\rm upper}$=35 K) emission.  These four transitions are representative of the range of morphologies seen in our ALMA data for lines that exhibit extended emission, including lines detected within the two wideband spectral windows (spws; Table~\ref{tab:obs}).  
As illustrated by Figure~\ref{fig:lines_4panel}, no line detected within our ALMA spectral coverage--including N$_2$H$^+$ (3-2)--exhibits an emission morphology consistently similar to that of the millimetre continuum.  Of the newly-detected millimetre continuum sources, only MM8 and MM10 are associated with molecular line emission--N$_2$H$^+$, H$_2$CO, and DCN--that is morphologically similar to the dust continuum, and so likely to arise from the same physical volume.  These two exceptions also differ from the other newly-detected sources in forming an extended ridge or filament of millimetre continuum emission south of MM2.  The two phenomena may be related: close examination of the line cubes suggests that the molecular emission, particularly for DCN, may trace the filament as a whole rather than two distinct continuum cores.  
Interestingly, CH$_{3}$OH emission is notably absent from the MM8-MM10 filament, in contrast to the other three molecules discussed above.
The lack of compact molecular line emission associated with the new millimetre continuum sources means that we cannot, in general, measure their individual LSR velocities from our ALMA data.

For most of the lines in our ALMA tuning that exhibit extended emission, that emission 
is dominated by the massive outflow driven by MM1.
In addition to the H$_2$CO line shown in Figure~\ref{fig:lines_4panel}a, our ALMA spectral coverage includes four other low-excitation lines of H$_2$CO (two with E$_{\rm upper}$=82 K and two with E$_{\rm upper}$=141 K) and one of CH$_3$OH (E$_{\rm upper}$=64 K); these relatively low-excitation lines of outflow-tracing molecules   
all exhibit similar morphologies.  
The $9_{-1}-8_{0}$ CH$_{3}$OH line shown in Figure~\ref{fig:lines_4panel}b is expected to be a Class I maser \citep[in the 36 GHz series;][]{Voronkov2012}.  With ALMA, \citet{ALMA_278_maser} observe negative excitation temperatures (implying population inversion) towards bright CH$_{3}$OH $9_{-1}-8_{0}$ emission peaks in G34.43+0.24 MM3.  These 278 GHz CH$_3$OH masers originate in post-shocked gas associated with outflow-cloud interaction regions \citep{ALMA_278_maser}, akin to lower-frequency Class I CH$_3$OH masers \citep[e.g.][]{Plambeck1990,C09}.  As shown in Figure~\ref{fig:lines_4panel}, in G11.92$-$0.61 the morphology of the CH$_{3}$OH $9_{-1}-8_{0}$ emission is broadly similar to that of H$_2$CO 4$_{0,4}-$3$_{0,3}$, but the CH$_{3}$OH emission is less extended and exhibits brighter, more compact peaks.  Many of these compact peaks are spatially coincident with 44 GHz Class I CH$_3$OH masers (Figure~\ref{fig:lines_4panel}b) and are candidate 278 GHz masers, though our ALMA beam is too large to definitively establish masing based on brightness temperature: the maximum brightness temperature measured towards the line of masers associated with the blueshifted MM1 outflow lobe is T$_{\rm B}$=163 K, while the 278 GHz emission spot immediately west of MM9 has a peak T$_{\rm B}$=40 K.  Interestingly, the line of 44 GHz and candidate 278 GHz CH$_3$OH masers southwest of MM1 (associated with its blueshifted outflow lobe) is coincident with DCN and H$_2$CO emission in our ALMA observations, suggesting that these molecules may be tracing shocked gas in an interaction region between the outflow and the surrounding cloud.
Also interestingly, while DCN is undetected towards the vast majority of the newly detected millimetre continuum sources, compact DCN emission is detected $\sim$0\farcs6 south of MM3 and $\sim$3\arcsec\/ south of MM1: in both cases, the emission is spatially coincident with weak millimetre continuum emission that did not meet our threshold of 5$\times$ the local rms for a named source, and is strongest in the 34 km s$^{-1}$ channel.

In general, the ALMA H$_2$CO and CH$_3$OH emission are more spatially extended and less collimated (less "jet-like") than the high-velocity $^{12}$CO(3-2) seen with the SMA, consistent with H$_2$CO and CH$_3$OH highlighting shocked outflow-cloud interaction regions, and potentially outflow cavity walls.  
In this context,
it is worth noting that the ALMA data trace gas much closer to the $v_{\rm LSR}$ than the SMA $^{12}$CO observations.  
The high abundance of $^{12}$CO and the limited \emph{uv}-coverage attainable with eight antennas with the SMA mean that near the systemic velocity, low-velocity outflow emission is confused with (poorly-imaged) emission from the ambient cloud. 
For example, the integrated velocity ranges for the main MM1 outflow in Figure~\ref{fig:irac_CO}a \emph{begin} at
$|v-v_{LSR}|$=15 km s$^{-1}$, and extend to significantly higher velocities ($|v-v_{LSR}|>$50 km s$^{-1}$), while   
the maximum velocity of the extended H$_2$CO 4$_{0,4}-$3$_{0,3}$ emission associated with the MM1 outflow is $|v-v_{LSR}|$$\lesssim$ 15 km s$^{-1}$ in our ALMA data.  
We note that the extensive emission associated with the MM1 outflow makes it difficult to assess whether new millimetre continuum sources located along the outflow axis are themselves physically associated with H$_2$CO or CH$_3$OH emission.  MM5, MM11, MM14, MM15, and MM18 overlap regions of H$_2$CO and/or CH$_3$OH emission in the ALMA cubes, but confusion from the MM1 outflow precludes using this emission to, for example, measure their individual LSR velocities.
The clumpy H$_2$CO emission, 278 GHz CH$_3$OH spots, and 44 GHz Class I CH$_3$OH masers detected in the southeastern region of our ALMA mosaic (e.g.\ south of the main MM1 outflow, and north of MM12) indicate that shocked gas is also present in this area.  In our ALMA data, this emission is predominantly redshifted, as are the 44 GHz CH$_3$OH masers \citep[][see also \citealt{C11sma}]{C09}.  It is unclear whether this emission represents a low-velocity, wide-angle component of the MM1 outflow, or whether it has another driving source.

The most exciting result from our ALMA line observations, with regard to the newly-detected millimetre continuum sources,
is the identification of an outflow driven by MM12 in H$_2$CO and CH$_3$OH emission (Figure~\ref{fig:lines_4panel}a,b).  
This outflow was not detected in our SMA $^{12}$CO(3-2) observations, likely due to sensitivity limitations.  For single-pointing observations like our SMA data, the sensitivity falls off (e.g. noise increases) with distance from the pointing centre; as shown in Figure~\ref{fig:irac_CO}, MM12 lies well outside the FWHP point of the SMA primary beam.  As seen with ALMA, MM12's outflow is characterised by a weaker eastern lobe and a brighter western lobe (with respect to the position of the millimetre continuum source).  The dominant western lobe is detected in both H$_2$CO and CH$_3$OH emission, which extends from slightly blueshifted (assuming v$_{\rm LSR}$=35 km s$^{-1}$, as above) to moderately redshifted velocities, with a total velocity extent of $\sim$34-42 km s$^{-1}$ in H$_2$CO 4$_{0,4}-$3$_{0,3}$ (see also Section~\ref{sec:outflow_prop}).  H$_2$CO and CH$_3$OH emission are also detected in the vicinity of MM7, MM9, and MM13, including emission from the MM7/MM9 outflow observed in $^{12}$CO; as noted above, the 278 GHz CH$_3$OH emission associated with this outflow may be masing, and is coincident with a 44 GHz Class I maser.  The high-velocity MM6 outflow is most clearly seen in $^{12}$CO(3-2) emission (compare Fig.~\ref{fig:irac_CO}a and Fig.~\ref{fig:lines_4panel}), though weak H$_2$CO and CH$_3$OH emission are detected near the systemic velocity ($|v-v_{LSR}|\lesssim$1 km s$^{-1}$) coincident with the $^{12}$CO lobes.

\section{Discussion}
\label{sec:dis}

\begin{table}
	\begin{minipage}{0.99\columnwidth}
	\centering
	\caption{Derived Properties of New Millimetre Continuum Sources.}
	\setlength{\tabcolsep}{0.05in}
	\label{tab:derived_cont_prop}
	\begin{tabular}{lccccccc} 
	\hline
	Source & Size$^{a}$ & T$_b^b$ & $\tau_{\rm dust}$ & M$_{\rm gas}$ & N$_{\rm H_2}^c$ & n$_{\rm H_2}^c$ & t$_{\rm ff}^c$ \\
	 &  &  & &  & $\times$10$^{23}$  & $\times$10$^{6}$ & $\times$10$^{3}$ \rule{0pt}{3.5ex}\\
 	 & (AU $\times$ AU) & (K) & & (M$_{\odot}$) &  (cm$^{-2}$) &  (cm$^{-3}$) & (yrs) \rule{0pt}{3.0ex}\\
	\hline
	MM4 & 1240 $\times$ 810 & 4.2 & 0.24 & 1.0 & 25 & 250 & 1.9 \\
	MM5 & 4470 $\times$ 1560 & 1.3 & 0.07 & 2.0 & 6.9 & 26 & 6.0 \\
	MM6 & 2110 $\times$ 1590 & 1.6 & 0.09 & 1.2 & 9.0 & 49 & 4.4 \\
	MM7 & 5180 $\times$ 3500 & 0.7 & 0.03 & 2.8 & 3.7 & 8.7 & 10 \\
	MM8 & 6080 $\times$ 1590 & 0.9 & 0.05 & 2.0 & 4.9 & 16 & 7.7\\
	MM9 & 3300 $\times$ 2930 & 0.7 & 0.03 & 1.4 & 3.5 & 11 & 9.2\\
	MM10 & 6620 $\times$ 2850 & 0.6 & 0.03 & 2.4 & 3.0 & 7.0 & 12\\
	MM11 & 4650 $\times$ 1900 & 0.7 & 0.03 & 1.3 & 3.6 & 12 & 8.8 \\
	MM12 & 3490 $\times$ 2720 &  0.7 & 0.03 & 1.4 & 3.5 & 11 & 9.1 \\
	MM13 & 4090 $\times$ 2730 & 0.4 & 0.02 & 1.1 & 2.4 & 7.2 & 11 \\
	MM14 & 2830 $\times$ 1630 & 0.6 & 0.03 & 0.6 & 3.0 & 14 & 8.2 \\
	MM15 & 4490 $\times$ 3670 & 0.3 & 0.02 & 1.1 & 1.6 & 4.1 & 15\\
	MM16 & 6470 $\times$ 4290 & 0.3 & 0.01 & 1.6 & 1.4 & 2.7 & 19 \\
	MM17 & 6760 $\times$ 3090 & 0.3 & 0.01 & 1.2 & 1.3 & 2.9 & 18 \\
	MM18 & 4390 $\times$ 2690 & 0.2 & 0.01 & 0.6 & 1.2 & 3.6 & 16 \\
    MM19 & $<$1370 & 0.1 & 0.007 & 0.2 & $>$2.7 & $>$19 & $<$7.0 \\ 
	\hline
	\end{tabular}
        \begin{flushleft}
         \small{$a$: 1/e$^2$ diameter of fitted Gaussian (Table~\ref{tab:obs_cont_prop}) at D=3.37 kpc (i.e. full width at the 1/e$^2$ intensity level$\sim$1.7$\times$FWHM).  The radius used in calculating N$_{\rm H_2}$ and n$_{\rm H_2}$ is half the geometric mean of the listed major and minor axes.  For MM19, which does not have a reliable fitted size, the quoted size limit is the geometric mean of the synthesised beam; the radius used in calculating N$_{\rm H_2}$ and n$_{\rm H_2}$ is half this value. }\\
         \small{$b$: Rayleigh-Jeans brightness temperature calculated using the fitted integrated flux density and FWHM deconvolved source size from Table~\ref{tab:obs_cont_prop}.  For MM19, which does not have a reliable fitted size, T$_b$ is instead calculated from the peak intensity and synthesised beamsize.}\\
         \small{$c$: Assuming spherical symmetry.  The free-fall time t$_{\rm ff}$ is calculated from the core density.}
        \end{flushleft}
	\end{minipage}
\end{table}

\subsection{Low-mass cores in a massive protocluster}
\label{sec:dis_phys_prop}

We estimate the physical properties of the newly-detected millimetre sources from their ALMA 1.05 mm continuum emission using the equation
\begin{equation}
    M_{gas} ({\rm in}~M_{\odot}) = \frac{4.79\times10^{-14}~ R ~S_{\nu}(Jy)~ D^2(kpc)~ C_{\tau_{dust}}}{B(\nu,T_{dust})~ \kappa_{\nu}},
\label{eq:dust}    
\end{equation}
which assumes isothermal dust emission and includes a correction for dust opacity, \begin{math} C_{\tau_{dust}} = \tau_{dust}/(1-e^{-\tau_{dust}}) \end{math}, where the dust opacity is estimated as \begin{math} \tau_{\rm dust} = -ln(1-\frac{T_b}{T_{\rm dust}}) \end{math}.  In equation~\ref{eq:dust}, R is the gas-to-dust mass ratio (assumed to be 100), $S_{\nu}$ is the integrated flux density from Table~\ref{tab:obs_cont_prop}, D is the distance to the source (3.37 kpc, Section~\ref{sec:intro}), and $B(\nu,T_{dust})$ is the Planck function.  For the newly detected sources MM4...MM19, we adopt $\kappa_{\rm 1.05mm}=$1.45 cm$^2$ g$^{-1}$, for grains with thick ice mantles in regions of high gas density \citep[interpolated from the values in column 9 of Table 1 of][]{OH94}, and T$_{\rm dust}=$20 K \citep[informed by NH$_{3}$ temperature measurements for an IRDC surrounding the GLIMPSE EGO G35.03+0.35;][]{Brogan2011}.
Derived source properties are presented in Table~\ref{tab:derived_cont_prop}, including T$_b$, $M_{gas}$, $\tau_{dust}$,
and H$_2$ column and number densities.  The latter two quantities are estimated from $M_{gas}$ and the source size assuming spherical geometry and using a mean molecular weight per hydrogen molecule $\mu_{H_2}$=2.8 \citep{Kauffmann2008}.  We note that the physical sizes presented in Table~\ref{tab:derived_cont_prop}, and used in calculating N$_{H_2}$ and n$_{H_2}$, are the 1/e$^2$ diameters of the fitted Gaussians (i.e., the full width at the 1/e$^2$ intensity level).  We use this heuristic in order to obtain densities that represent the values in the bulk of the core volume rather than in the central region.

The VLA nondetections of MM4...MM19 at 3 and 0.9 cm (Section~\ref{sec:cont_results}) support the assumption that the 1.05 mm emission from these sources arises from thermal dust emission.  If the millimetre-wavelength emission was attributable to either optically thin free-free emission ($\alpha=-0.1$) or free-free emission from an ionised wind or jet ($\alpha\sim0.6$), these sources would be readily detected in our VLA images (S/N$>$30 at 3 cm for $\alpha=0.6$).  Even in the extreme case of an ionised source that remained optically thick from 0.9 cm up to 1.05 mm (e.g.\ $\alpha=$2), the 4$\sigma$ VLA 0.9 cm limit would correspond to S$_{\rm 1.05 mm}\sim$2.3 mJy; of the newly detected sources, only MM19 has an integrated flux density below this level.  In contrast, spectral indices $\alpha\sim$3-4, as expected for dust emission, are consistent with the VLA nondetections.

The derived physical properties of the newly-detected millimetre sources in G11.92$-$0.61 are broadly similar to those of low-mass cores in nearby star-forming regions.  The gas masses of the newly-detected cores (for our assumed dust temperature and opacity) range from 0.2-2.8 M$_{\odot}$, with a median mass of 1.3 M$_{\odot}$ and a median radius of $\sim$1600 au.
In Serpens, Perseus, and Ophiuchus, low-mass prestellar and protostellar cores typically have masses of $\sim$0.1-10 M$_{\odot}$ and sizes of several thousand AU \citep[e.g.][]{Hogerheijde1999,Kirk2006,Enoch2008,Simpson2008}.  
In contrast, discs around low-mass (proto)stars generally have size scales $<$ 500 AU and masses $<$ a few tenths M$_{\odot}$, though younger objects may have more massive discs \citep[e.g.][]{Andrews2010,Enoch2011}. 
The inferred masses and sizes of MM4..MM18, as compared to the properties of low-mass core populations in well-studied nearby regions, lead us to conclude that the newly-discovered ALMA sources in G11.92$-$0.61 are most likely to be low-mass prestellar and/or protostellar cores.  As discussed above and in Section~\ref{sec:cont_results}, it is possible that MM19--which has a point-source MIR counterpart, and for which we cannot rule out optically thick free-free as an explanation for its millimetre emission--may differ in nature from the other new ALMA sources, but additional deep multi-wavelength observations would be required to investigate this possibility.
We note that the newly detected cores in G11.92$-$0.61 do have somewhat higher source-averaged densities (Table~\ref{tab:derived_cont_prop}; median n$_{\rm H_2}\sim$1$\times$10$^{7}$ cm$^{-3}$) than the prestellar and protostellar cores of Serpens, Perseus, and Ophiuchus: \citet{Enoch2008} found average mean densities of a few $\times$ 10$^{5}$ cm$^{-3}$, with the tail of the distribution extending out to $\sim$5$\times$10$^{6}$ cm$^{-3}$.  However, the densities from \citet{Enoch2008} are also averaged over a larger sizescale, of 10,000 au.

\begin{figure}
	\includegraphics[width=\columnwidth]{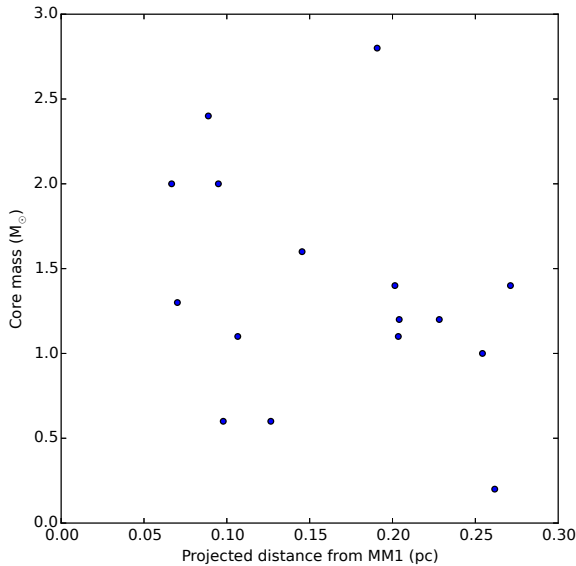}
    \caption{Core mass (from Table~\ref{tab:derived_cont_prop}) vs. projected distance from MM1 for the newly-detected millimetre sources MM4..MM19; the median offset from MM1 is $\sim$0.17 pc (Section~\ref{sec:dis_phys_prop}).   }
    \label{fig:mass_v_sep}
\end{figure}

The lack of single-dish bolometer observations at our ALMA observing frequency precludes a straightforward calculation of the percentage of the total flux density recovered with ALMA.  The alternative, calculating the percentage of 
the clump \emph{mass} attributable to compact cores, entails 
considerable uncertainties due to the need to assume opacities and dust temperatures at both the core and the clump 
scale.  In addition, both the ATLASGAL and JCMT SCUBA 850 $\mu$m bandpasses include the $^{12}$CO(3-2) line, which can contribute significantly 
to the total 
flux density measured with (sub)millimetre bolometers in regions with outflows \citep[e.g. $\sim$20-30\% for SCUBA
850$\mu$m;][]{Hatchell2009}.  We measure an integrated flux density of 11.0$\pm$0.3 Jy from the ATLASGAL 870 $\mu$m
map\footnote{Obtained from the ATLASGAL database server at \\
http://atlasgal.mpifr-bonn.mpg.de/cgi-bin/ATLASGAL\_DATABASE.cgi}  
\citep{Schuller2009} within the 30\% power level of our ALMA mosaic.
(This value is intermediate between those in the two ATLASGAL catalogs: 9.84 Jy in the \citealt{Csengeri2014} Gaussclumps catalog, optimised 
for small-scale embedded structures, and  38.83 Jy in the \citealt{Contreras2013} SExtractor catalog optimised for larger-scale clump and cloud structures.)  For $\kappa_{\rm 870 \mu m}=$ 1.85 cm$^{2}$ g$^{-1}$ \citep[as in][]{Schuller2009} and T$_{\rm dust}=$20 K, and neglecting line contamination, the measured ATLASGAL flux density equates to a mass reservoir of $\sim$700 M$_{\odot}$ within the $\sim$0.7 pc diameter of our ALMA mosaic.  The total mass of the newly-detected ALMA sources MM4..MM19 is $\sim$22 M$_{\odot}$, or $\sim$3\% of the ATLASGAL clump mass (these values increase to $\sim$33 M$_{\odot}$ and $\sim$5\% if we instead assume T$_{\rm dust}$=15 K, closer to the NH$_3$ temperatures measured in quiescent regions of the IRDC G11.11-0.12 P6 by \citealt{Wang2014}).  Estimates of the total mass in compact cores---e.g.\ including the previously-known sources MM1, MM2, and MM3--depend on the mass estimates for 
these sources, which are quite sensitive to the opacity correction term in equation~\ref{eq:dust} \citep[in particular for MM2, see also Table 2 of][]{Cyganowski2014}.  While a detailed analysis of the massive sources is beyond the scope of this paper \citep[see also][]{Ilee2016,Cyganowski2014}, we conservatively estimate that MM1, MM2, and MM3 together account for $\sim$5-10\% of the ATLASGAL clump mass.  It is interesting that, at sub-arcsecond resolution, the fraction of the clump mass resolved into high-mass cores is of the same order of magnitude as that resolved into low-mass cores (indeed, within a factor of $\sim$2-3), and that all compact cores detected to date account for $\lesssim$15\% of the total clump-scale mass reservoir.  There may, of course, be additional low-mass cores below the sensitivity limit of our ALMA observations; the rms noise level varies significantly across the map (Section~\ref{sec:obs_alma}), corresponding to a 5$\sigma$ mass sensitivity for T$_{\rm dust}$=20 K that ranges from $\sim$0.1 M$_{\odot}$ (in the vicinity of MM19) to $\sim$0.3 M$_{\odot}$ near the edge of the mosaic, and in the most dynamic-range-limited regions near the high-mass cores (the corresponding 5$\sigma$ mass sensitivities are $\sim$0.2-0.4 M$_{\odot}$ for T$_{\rm dust}$=15 K, and $\sim$0.4-0.8 M$_{\odot}$ for T$_{\rm dust}$=10 K).

\begin{figure}
	\includegraphics[width=\columnwidth]{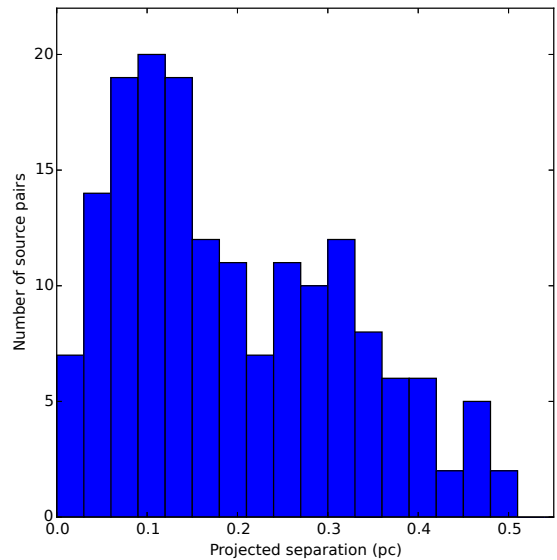}
    \caption{Distribution of projected separations for all unique millimetre source pairs in G11.92$-$0.61, including the massive protocluster members MM1..MM3.}
    \label{fig:separation_hist}
\end{figure}

Notably, the newly-detected ALMA sources are, as a population, found significantly offset from the cluster centre, in the outer reaches of the accretion reservoir.  Figure~\ref{fig:mass_v_sep} shows core mass plotted against projected distance from MM1 for the low-mass cores MM4..MM19, while Figure~\ref{fig:separation_hist} presents the distribution of projected separations for \emph{all} unique millimetre source pairs (e.g.\ including the massive protocluster members MM1..MM3 as well as MM4..MM19). 
As illustrated in Figure~\ref{fig:mass_v_sep}, the median offset from MM1 of the low-mass cores is $\sim$0.17 pc.  While this could in part be attributable to the higher rms noise near the massive cores, we note that the outflow-driving sources MM6, MM7/9, and MM12 are all 0.2-0.3 pc from MM1, as is MM4, the highest-density and most compact of the low-mass cores.  
In the literature, a radius of 0.1 pc is typically adopted for massive cores \citep[e.g.][and references therein]{Tan2014,Myers2013}.  In this context, it is interesting to note that the distribution of source pair separations shown in Figure~\ref{fig:separation_hist}--which includes the massive sources MM1..MM3--exhibits two peaks, at $\sim$0.3 pc and $\sim$0.1 pc.  Five low-mass cores are also located $<$0.1 pc (in projection) from MM1 (Figure~\ref{fig:mass_v_sep}), e.g.\ within the expected sizescale of a high-mass core.    
Both results are consistent with theoretical predictions from competitive-accretion-style cluster formation models, which predict low-mass cores and star formation both near the most massive cluster member (within $<$0.1 pc) and throughout an accretion reservoir with a scale of a few tenths of a parsec \citep[e.g.][see also Section~\ref{sec:intro}]{Smith2009,Wang2010}.

\begin{table*}
	\begin{minipage}{0.9\textwidth}
	\centering
	\caption{Outflow Properties: $^{12}$CO.}
	\label{tab:outflow_co}
	\begin{tabular}{lccccc} 
	\hline
	& $^{12}$CO(3-2) & $^{12}$CO(3-2) & $^{12}$CO(3-2) & $^{12}$CO(2-1)$^{a}$ & $^{12}$CO(2-1)$^{a}$ \\
	Parameter$^{b}$ & MM6 Outflow & MM7/9 Outflow$^{c}$ & MM1 Outflow & MM6 Outflow & MM1 Outflow \\
	\hline
	v[red] (km s$^{-1}$) & 48.5 -- 84.5  & 42.5 -- 56.0 & 50.0 -- 74.0 & 48.1 -- 81.1 & 48.1 -- 71.2\\
	v[blue] (km s$^{-1}$) & 6.5 -- 27.5 & 23.0 -- 30.5 & $-$17.5 -- 20.0 & 5.3 -- 21.8 & $-$24.4 -- 21.8  \\
	M$_{\rm out}$-total (M$_{\odot}$) & 0.02 & 0.03 & 0.10 & 0.04 & 0.6 \rule{0pt}{3.5ex}\\
	P$_{\rm out}$-total (M$_{\odot}$ km s$^{-1}$) & 0.3 & 0.3 & 2.3 & 0.8 & 12 \rule{0pt}{3.5ex}\\
    E$_{\rm out}$-total (ergs) & 7.9$\times$10$^{43}$ & 3.2$\times$10$^{43}$ & 6.0$\times$10$^{44}$ & 1.9$\times$10$^{44}$ & 3.0$\times$10$^{45}$  \rule{0pt}{3.5ex}\\
    Length[red] (pc) & 0.04 & 0.04 & 0.13 & 0.07 & 0.2 \rule{0pt}{3.5ex}\\
    Length[blue] (pc) & 0.05 & 0.04 & 0.23 & 0.07 & 0.3 \\
    t$_{\rm dyn}$[red] (yrs) & 800 & 2000 & 3300 & 1400& 4400 \rule{0pt}{3.5ex}\\
    t$_{\rm dyn}$[blue] (yrs) & 1800 & 3000 & 4300 & 2300 & 5300 \\
    {\.M}$_{\rm out}$-total (M$_{\odot}$ yr$^{-1}$)  & 1.3$\times$10$^{-5}$ & 1.2$\times$10$^{-5}$ & 2.6$\times$10$^{-5}$  &   2.5$\times$10$^{-5}$ & 1.2$\times$10$^{-4}$ \rule{0pt}{3.5ex}\\
    {\.P}$_{\rm out}$-total (M$_{\odot}$ km s$^{-1}$ yr$^{-1}$)  &  3.0$\times$10$^{-4}$ & 1.3$\times$10$^{-4}$ & 5.9$\times$10$^{-4}$ & 5.1$\times$10$^{-4}$ & 2.4$\times$10$^{-3}$\rule{0pt}{3.5ex}\\
    L$_{\rm mech}$-total (L$_{\odot}$)  & 0.7 & 0.1 & 1.3 & 1.0 & 5\rule{0pt}{3.5ex}\\
    \hline
	\end{tabular}
        \begin{flushleft}
          \small{$a$: MM1 outflow: from \citet{C11sma}, rescaled for D=3.37 kpc.  MM6 outflow: properties estimated from the $^{12}$CO(2-1) data presented in \citet{C11sma}.}\\  
          \small{$b$: "Total" quantities are the sum of the red and blue lobes.}\\
            \small{$c$: Lengths measured from MM9.}\\
        \end{flushleft}
	\end{minipage}
\end{table*}

\subsection{Outflow properties}
\label{sec:outflow_prop}

Molecular outflows are driven by protostars of all masses during their formation processes, and so are ubiquitous in star-forming regions \citep[e.g.][]{Richer2000,Zhang2001,Arce2007,Frank2014}.  The observational evidence for correlations, over many orders of magnitude, between outflow properties and those of the driving source--primarily from large-scale surveys of protostellar outflows with single-dish telescopes \citep[e.g.][]{Cabrit1992,Shepherd1996,Beuther2002,Wu2004,Maud2015out}--suggests that measuring outflow properties may provide an additional means of investigating the nature of the driving protostar.

To this end, we estimate the properties of the outflows driven by MM6, MM7/9, and MM12 from the SMA $^{12}$CO(3-2) and the ALMA H$_2$CO 4$_{0,4}-$3$_{0,3}$ data, using an approach similar to that of \citet{C11sma}.  For each outflow lobe, we define a mask based on the 3$\sigma$ contour in the integrated intensity image (moment 0 map) of that lobe.  The emission within this mask is then measured in each velocity channel that contributes to the moment 0 map, and the gas mass (assuming optically thin emission) calculated using \citep[following equation B2 of][]{MangumShirley2015} 
\begin{equation}
    M_{\rm out} = \frac{4.788\times3c^2\times1.36 m_{H_2}\times Q(T_{ex}) e^\frac{E_{\rm upper}}{T_{ex}} D^2 \int S_\nu dv}{16\pi^3\nu^3 \mu^2 S \chi}
\end{equation}
where M$_{\rm out}$ is the gas mass (in M$_{\odot}$), Q(T$_{ex}$) is the partition function, T$_{ex}$ is the (assumed) excitation temperature (in K), D is the source distance (in kpc), S$_\nu$ is the line flux density measured as described above (in Jy), $dv$ is the velocity channel width in km s$^{-1}$,  $\chi$ is the (assumed) abundance of the observed molecule with respect to H$_2$, and a mean gas atomic weight of 1.36 is assumed \citep{Qiu2009,C11sma}.
The upper energy level (E$_{\rm upper}$, in K), frequency ($\nu$, in GHz), and line strength ($\mu^2$S, in debye$^{2}$) are properties of the observed molecular transition.  
We adopt T$_{ex}$=30 K, $\chi_{\rm CO}$=10$^{-4}$ \citep[e.g.][and references therein]{Dunham2014}, and $\chi_{\rm H_{2}CO}$=10$^{-8}$, noting that since shocks enhance the H$_2$CO abundance, it varies as a function of velocity and of position within individual outflows (e.g. \citealt{Tafalla2010}, \citealt{vDB98}; \citealt{Bachiller1997}, for example, find abundances from 4$\times$10$^{-9}$ to 6$\times$10$^{-7}$ in the L1157 outflow).
From the Cologne Database for Molecular Spectroscopy (CDMS) entries in the Splatalogue\footnote{http://www.cv.nrao.edu/php/splat/} spectral line database \citep{Remijan2010}, $\mu^2S=$0.03631 debye$^{2}$ and  $\mu^2S=$21.74159 debye$^{2}$ for $^{12}$CO(3-2) and H$_2$CO 4$_{0,4}-$3$_{0,3}$, respectively; interpolating from the partition function values in the CDMS \citep{Muller2001,Muller2005}, we use $Q(30 K)=11.19$ for $^{12}$CO and $Q(30 K)=95.1$ for H$_2$CO.

Tables~\ref{tab:outflow_co} and ~\ref{tab:outflow_h2co} present estimates for the outflow properties derived from the SMA $^{12}$CO(3-2) and the ALMA H$_2$CO 4$_{0,4}-$3$_{0,3}$ data, respectively.  The reported mass is the sum of the mass in all contributing velocity channels (listed as v[red] and v[blue] for redshifted and blueshifted lobes, respectively) and the momentum and energy are calculated \citep[following e.g.][]{C11sma,Qiu2009} as
\begin{equation}
    P_{\rm out}= \sum\limits_i M^i_{\rm out} \Delta v
\end{equation}
and
\begin{equation}
    E_{\rm out} = \frac{1}{2} \sum\limits_i M^i_{\rm out} (\Delta v)^2
\end{equation}
where $M^i_{\rm out}$ is the mass in a given velocity channel and \begin{math} \Delta v = |v_{\rm channel}-v_{\rm LSR}| \end{math}.  In the absence of compact molecular line emission associated with MM6, MM7, MM9, and MM12 with which to measure their LSR velocities (Section~\ref{sec:line_results}), we adopt $v_{\rm lsr}$=35 km s$^{-1}$ for all of our outflow analysis.  For each outflow lobe, we estimate the dynamical timescale as \begin{math} t_{\rm dyn}=L_{\rm outflow}/v_{\rm max} \end{math}, where L$_{\rm outflow}$ is the outflow length (measured from the driving millimetre continuum source) and $v_{\rm max}$ is the maximum velocity of the blue- or red-shifted outflow gas.  Using these dynamical timescales, we then estimate the timescale-dependent properties \.M$_{\rm out}$ (mass outflow rate, M$_{\rm out}$/t$_{\rm dyn}$),  \.P$_{\rm out}$ (momentum outflow rate, P$_{\rm out}$/t$_{\rm dyn}$), and L$_{\rm mech}$ (mechanical luminosity, E$_{\rm out}$/t$_{\rm dyn}$).  
Rather than potentially overcorrect our estimates of outflow dynamical properties \citep[e.g.][]{Downes2007,Dunham2014}, the values in Tables~\ref{tab:outflow_co} and ~\ref{tab:outflow_h2co} are not corrected for inclination. 
For all properties listed in Tables~\ref{tab:outflow_co} and ~\ref{tab:outflow_h2co}, the outflow "total" is the sum of estimates for the red- and blue-shifted outflow lobes.

We emphasise that our estimates of outflow properties--in particular for the larger-scale MM1 outflow--are affected by the spatial filtering of the interferometers, the confusion of outflow with cloud emission near the systemic velocity (which dictates the velocity ranges over which outflow properties can be integrated), and the assumption of optically thin emission.  
All of these factors act to make our estimated outflow masses underestimates \citep[see also discussion in][]{C11sma,Dunham2014}.  
Additionally, in a study of outflows driven by isolated low-mass protostars, \citet{Dunham2014} found that for a given outflow, the mass and dynamical properties calculated from $^{12}$CO(3-2) emission were, on average, an order of magnitude lower than if the same properties were calculated from $^{12}$CO(2-1).  To investigate this effect in our data, Table~\ref{tab:outflow_co} also presents the estimated properties of the MM1 outflow as derived from $^{12}$CO(2-1) SMA observations \citep[$\sim$2\farcs4 resolution;][rescaled to D=3.37 kpc]{C11sma}.  In addition, we reanalysed the $^{12}$CO(2-1) data to extract estimated properties for the MM6 outflow, which was detected \citep[e.g. Fig. 6a of][]{C11sma} but not resolved in these observations.  Comparison of the estimates presented in Table~\ref{tab:outflow_co} shows that the differences in the outflow properties estimated from the 2\farcs4 resolution $^{12}$CO(2-1) and the 0\farcs76 resolution $^{12}$CO(3-2) data are greatest for the MM1 outflow (with the $^{12}$CO(2-1) estimates being 6$\times$, 4.6$\times$, and 4$\times$ greater for M$_{\rm out}$, \.M$_{\rm out}$, and \.P$_{\rm out}$, respectively).  This is as expected: compared to the outflows from MM6 and MM7/MM9, the MM1 outflow is significantly more spatially extended (and so susceptible to spatial filtering), and is confused with cloud emission over a wider velocity range around the v$_{\rm LSR}$ \citep[leading to missing more of the mass, momentum, and energy associated with outflow gas at near-systemic velocities, see e.g.][]{Dunham2014,Maury2009,Offner2011,vanderMarel2013}.  In contrast, the $^{12}$CO(2-1) and $^{12}$CO(3-2)-derived M$_{\rm out}$, \.M$_{\rm out}$, and \.P$_{\rm out}$ for the MM6 outflow differ by a factor of $\lesssim$2.  

\begin{table}
	\begin{minipage}{0.95\columnwidth}
	\caption{Outflow Properties: H$_{2}$CO.}
	\label{tab:outflow_h2co}
	\setlength{\tabcolsep}{0.04in}
	\begin{tabular}{lcc} 
	\hline
	Parameter$^{a}$ & MM12 Outflow & MM7/9 Outflow$^{b}$  \\
	\hline
		\hline
	v[red] (km s$^{-1}$) & 36.0 -- 42.0 & 36.5 -- 39.0 \\
	v[blue] (km s$^{-1}$) & \nodata &  30.0-33.5  \\
	M$_{\rm out}$-total (M$_{\odot}$) & 0.09 & 0.05 \rule{0pt}{3.5ex}\\
	P$_{\rm out}$-total (M$_{\odot}$ km s$^{-1}$) & 0.3 & 0.12 \rule{0pt}{3.5ex}\\
    E$_{\rm out}$-total (ergs) & 1.3$\times$10$^{43}$ &  3.6$\times$10$^{42}$ \rule{0pt}{3.5ex}\\
    Length[red] (pc) & 0.05 & 0.01 \rule{0pt}{3.5ex}\\
    Length[blue] (pc) & \nodata & 0.03 \\
    t$_{\rm dyn}$[red] (yrs) & 5800 & 3200 \rule{0pt}{3.5ex}\\
    t$_{\rm dyn}$[blue] (yrs) & \nodata & 5400 \\
    {\.M}$_{\rm out}$-total (M$_{\odot}$ yr$^{-1}$)  & 1.5$\times$10$^{-5}$ &  1.2$\times$10$^{-5}$ \rule{0pt}{3.5ex}\\
    {\.P}$_{\rm out}$-total (M$_{\odot}$ km s$^{-1}$ yr$^{-1}$)  & 5.3$\times$10$^{-5}$ &  3.0$\times$10$^{-5}$\rule{0pt}{3.5ex}\\
    L$_{\rm mech}$-total (L$_{\odot}$)  & 0.02 &  0.007 \rule{0pt}{3.5ex}\\
    \hline
	\end{tabular}
        \begin{flushleft}
          \small{$a$: "Total" quantities are the sum of the red and blue lobes.}\\
            \small{$b$: Lengths measured from MM9.}\\
        \end{flushleft}
	\end{minipage}
\end{table}

The fact that the MM7/MM9 outflow is detected in both $^{12}$CO(3-2) and H$_2$CO 4$_{0,4}-$3$_{0,3}$ provides a means of assessing whether these two tracers give comparable estimates of outflow properties for compact low-mass flows in G11.92$-$0.61.  (We do not use H$_2$CO to calculate properties for the MM1 outflow due to the morphological and kinematic evidence that, for this massive outflow, the H$_2$CO emission is dominated by shocked interface regions rather than the high-velocity jet seen in $^{12}$CO; Section~\ref{sec:line_results}.)  Comparing the estimates of the MM7/9 outflow properties derived from $^{12}$CO(3-2) (Table~\ref{tab:outflow_co}) and H$_2$CO 4$_{0,4}-$3$_{0,3}$ (Table~\ref{tab:outflow_h2co}), we find remarkably good agreement (within a factor of 2) for the total outflow mass and mass outflow rate, considering the uncertainty in the H$_2$CO abundance and the different velocity ranges probed by the two species.  For \.P$_{\rm out}$ and L$_{\rm mech}$, the H$_2$CO-based estimates are considerably lower, consistent with the stronger dependence of these properties on the velocity of the outflowing gas ($\Delta v$, as defined above) and with H$_2$CO tracing much more moderate velocities (e.g.\ closer to the $v_{\rm LSR}$) than $^{12}$CO.  

Considering the (large) uncertainties associated with different resolutions, spatial filtering, and molecular tracers, the properties we estimate for the MM6, MM7/9, and MM12 outflows are reasonably consistent with the ranges found in surveys of outflows from low-mass (proto)stars \citep[e.g.][]{Bontemps1996,Arce2006,Dunham2014}.  More tellingly, the estimated properties of the MM6, MM7/9, and MM12 outflows are fairly similar to each other, but distinct from those of the massive MM1 outflow.  The MM6, MM7/9, and MM12 outflows have outflow masses of a few hundredths M$_{\odot}$, momenta of a few tenths M$_{\odot}$ km s$^{-1}$, lengthscales of $\lesssim$0.05 pc (for each lobe), mass outflow rates of $\sim$10$^{-5}$ M$_{\odot}$ yr$^{-1}$, and momentum outflow rates of a few $\times$10$^{-5}$ to a few $\times$10$^{-4}$ M$_{\odot}$ km s$^{-1}$ yr$^{-1}$.  Comparing the properties of the MM6 and MM1 outflows derived from $^{12}$CO(2-1)--of the available datasets, the one in which the properties of the MM1 outflow will be least severely underestimated--the size (length), mass, momentum, and energy of the MM1 outflow are larger by more than an order of magnitude, and \.M$_{\rm out}$, \.P$_{\rm out}$, and L$_{\rm mech}$ are larger by factors of $\sim$5.  As discussed above, the MM1 outflow is expected to be most affected by all of the factors that contribute to underestimating outflow properties, so the true differential between the MM1 outflow and those from MM6, MM7/9, and MM12 is likely even greater.  

Without well-sampled spectral energy distributions (SEDs), we cannot estimate the luminosities of the individual members of the G11.92$-$0.61 protocluster, and so cannot place MM6, MM7/9, and MM12 in the \.P$_{\rm out}$-L$_{bol}$ space often used to compare outflows from low- and high-mass protostars \citep[e.g.][]{Beuther2002,Maud2015out,Cunningham2016}.  
Interestingly, combining a sample of high-mass protostars in Cygnus X and low-mass Class 0 and I sources from \citet{Bontemps1996}, \citet{DuarteCabral2013} report an analogous correlation between \.P$_{\rm out}$ and M$_{\rm env}$ (e.g. their Figure 4).  (Note that M$_{\rm env}$ as measured by \citealt{DuarteCabral2013} is the mass on $\sim$4000 au scales around individual protostars, as distinct from the clump-scale masses from single-dish observations considered by \citealt{Beuther2002core}, \citealt{Maud2015core} and \citealt{deVilliers2014}.)  Notably, there is substantial overlap between the \citet{DuarteCabral2013} high-mass sources and the \citet{Bontemps1996} low-mass Class 0 sources on the momentum flux axis, encompassing the range of values we measure for the MM6, MM7/9, and MM12 outflows.  In contrast, there is a clear separation between the two populations in envelope mass.  
While our ALMA core mass estimates are not directly comparable to the envelope masses from \citet{DuarteCabral2013} (which are derived from SED fits), it is interesting to note that if
the core masses from Table~\ref{tab:derived_cont_prop} are used to place the MM6, MM7/9, and MM12 outflows on the \citet{DuarteCabral2013} \.P$_{\rm out}$-M$_{\rm env}$ plot, these sources would fall within the region of parameter space occupied by low-mass Class 0 sources.  Taken together, the core mass estimates from Section~\ref{sec:dis_phys_prop}, the relative properties of the G11.92$-$0.61 outflows (discussed above), and the comparison with samples from the literature all tend to the same conclusion: the outflow-driving newly-detected millimetre sources in G11.92$-$0.61 are low-mass protostellar cores.

\subsection{MM3: Intermediate mass star formation in G11.92-0.61?}
\label{sec:dis_mm3}

The place occupied by MM3 in the hierarchy of low/intermediate/high mass star formation in G11.92$-$0.61 is unclear. The presence of a 6.7 GHz Class II CH$_{3}$OH maser and 8 $\mu$m and 24 $\mu$m emission argue for MM3 being a massive (proto)star \citep[Section~\ref{sec:intro},][]{C11sma}, but its millimetre continuum emission is significantly weaker than that from the massive sources MM1 and MM2 \citep[Table~\ref{tab:obs_cont_prop}, see also][]{C11sma}.
In our VLA 3 and 0.9 cm images, a point-source counterpart to MM3-C1 is detected: the fitted positions and integrated flux densities from two-dimensional Gaussian fits are $\alpha$=18$^{h}$13$^{m}$58$^{s}$.1345 $\delta$=$-$18$^{\circ}$54\arcmin16\farcs27 (J2000) and S$_{\rm int}$=52$\pm$9 $\mu$Jy at 3 cm and $\alpha$=18$^{h}$13$^{m}$58$^{s}$.1350 $\delta$=$-$18$^{\circ}$54\arcmin16\farcs258 (J2000) and S$_{\rm int}$=149$\pm$11 $\mu$Jy at 0.9 cm.  The cm-wavelength spectral index $\alpha_{\rm 3.0-0.9cm}$=0.9$\pm$0.4, while the spectral index between 0.9 cm and 1.05 mm is 2.1$\pm$0.4; these measurements raise the possibility that much or all of the 1.05 mm flux density of MM3-C1 may be attributable to optically-thick free-free emission, particularly if multiple ionised gas components (e.g., an ionised jet and a hypercompact HII region) are present \citep[see e.g. the discussion of MM1's centimetre-wavelength emission in][]{Ilee2016}.  

\citet{C11sma} report tentative evidence for an outflow driven by MM3 (the "northern outflow" in their Table 7); the properties of this outflow (as derived from $^{12}$CO(2-1)) are intermediate between those of the MM1 and MM6 outflows, with the red lobe also having a dynamical timescale longer than those of the other outflows in the region ($\gtrsim$10,000 years).  We also see some evidence for an outflow driven by MM3 in our ALMA H$_{2}$CO data, but at the comparatively low velocities probed by this tracer, the possible MM3 outflow is too confused with the main MM1 outflow to allow reliable estimates of physical properties to be made.  
The candidate MM3 outflow also shares some characteristics--such as its prominence at relatively moderate velocities--with the "relic" outflows observed in association with some ultracompact (UC) HII regions \citep[e.g. in G18.67+0.03 and G5.89$-$0.39;][]{Cyganowski2012,Hunter2008}.  These outflows are thought to no longer be actively driven, but to have been driven by the now-ionising source prior to the formation of a UC HII region. 
In summary, from the available evidence, it remains difficult to establish the evolutionary state of MM3-C1, and whether it is a high-mass or an intermediate-mass (proto)star.

\subsection{Simultaneous low- and high-mass star formation}
\label{sec:dis_simul}

The fact that we detect molecular outflows from both high-mass (MM1) and low-mass (MM6, MM7/9, and MM12) sources within the G11.92$-$0.61 protocluster is strong evidence that high-mass and low-mass stars are forming simultaneously, e.g.\ are approaching their final masses through concurrent ongoing accretion.  While the exact mechanism(s) continue to be subjects of study, there is wide consensus that outflow is intrinsically linked to accretion \citep[e.g.][and references therein]{Frank2014}; therefore, the presence of an active outflow implies ongoing accretion.  
The outflows detected in G11.92$-$0.61--both from MM1 and from the low-mass cores--all have roughly comparable, and short, dynamical timescales, of $\lesssim$6000 years.
The free-fall timescales for the low-mass cores ($\sim$2$\times$10$^3$-2$\times$10$^4$ years; Table~\ref{tab:derived_cont_prop}) are also consistent with ongoing low-mass star formation; even if some of these cores are currently starless, the brief free-fall times imply that star formation could be imminent.
The presence of the massive starless core candidate MM2 within the same protocluster also argues for in-progress massive star formation; while there is no direct evidence that MM2 is accreting (due to the lack of associated line emission detected to date), the free-fall time implied by the high density of this massive core is extremely short, $\lesssim$1000 years \citep{Cyganowski2014}.
We note that observations of a single protocluster cannot speak directly to the question of whether accretion \emph{began} at the same moment for the high- and low-mass protostars, e.g. whether their formation timescales have the same absolute t=0.  Addressing this question will require observations similar to the ALMA data presented here for an ensemble of targets of different ages, selected to sample the time-sequence of protocluster formation.  

It is also interesting to note that the inferred masses for our newly-detected ALMA millimetre cores (Section~\ref{sec:dis_phys_prop}) are concentrated in the range of 1-2 M$_{\odot}$.  We do not detect a larger number of $<$1 M$_{\odot}$ cores, as would be expected if the distribution of core masses followed the shape of the stellar initial mass function (IMF) \citep[e.g.][see Section ~\ref{sec:dis_phys_prop} for discussion of our ALMA sensitivity limits]{Rathborne2009}.  There is also an apparent lack of \emph{cores} with masses intermediate between $\sim$3 M$_{\odot}$ and the massive starless core candidate MM2 \citep[M$\gtrsim$30 M$_{\odot}$;][]{Cyganowski2014}.  We emphasise, however, that the masses derived from our ALMA observations include only the circumprotostellar material, and \emph{not} the masses of protostars that have already formed--such as those driving the MM6, MM7/9, and MM12 outflows.    

Comparing our observational results with theoretical model predictions, in detail, will require synthetic observations of the models, to allow direct comparisons of observable diagnostics.  In interferometric observations like our ALMA images, what is detected as a ``core'' is expected to depend on a convolution of spatial filtering, column density, and dust temperature, with the dust temperature depending in turn on internal and external heating \citep[see also][]{Smith2009}.  Only by creating synthetic observations that share the key characteristics of observed data (e.g.\ angular resolution, spatial filtering, noise) can we establish what observable core population would be expected in dynamic scenarios of cluster formation, where accretion proceeds across a variety of size scales and along non-axisymmetric structures such as filaments \citep[e.g.][]{Smith2016}.

\section{Conclusions}
\label{sec:conclusions}

Our sub-arcsecond ($\sim$1400 au) resolution ALMA Cycle 2 1.05 mm observations of G11.92$-$0.61 reveal a rich population of low-mass cores surrounding the central three massive members (MM1..MM3) of this (proto)cluster.  Our main findings are:

\begin{itemize}

    \item We detect 16 new millimetre continuum sources within the $\sim$0.7 pc diameter of our ALMA mosaic.  All are undetected in deep VLA 3 cm and 0.9 cm images, consistent with the millimetre flux arising from thermal dust emission. \\
    
    \item The gas masses of the newly-detected sources, inferred from their dust emission, range from 0.2-2.8 M$_{\odot}$, with a median of 1.3 M$_{\odot}$ (for T$_{\rm dust}$=20 K).  For the population of newly detected sources, the median radius is $\sim$1600 au and the median H$_2$ number density is n$_{\rm H_2}\sim$1$\times$10$^{7}$ cm$^{-3}$, broadly consistent with low-mass prestellar and protostellar cores in nearby star-forming regions. \\ 
    
    \item The newly-detected low-mass cores are, on average, found in the outer reaches of the accretion reservoir: the median projected separation of a newly-detected source from the high-mass protostar MM1 is $\sim$0.17 pc.\\
    
    \item At least 3 of the newly-detected low-mass cores (MM6, MM7/9, and MM12) are driving molecular outflows, traced by high-velocity collimated $^{12}$CO(3-2) emission (observed with the SMA) and/or by H$_2$CO and CH$_3$OH emission observed with ALMA.  The outflows indicate that these low-mass protostars are currently accreting, as is the neighbouring high-mass protostar MM1.

\end{itemize}

Our discovery of low-mass cores within the accretion reservoir of forming massive stars is consistent with the predictions of competitive accretion-style cluster formation models, in which a cluster environment is required to form high-mass stars \citep[e.g.][]{BonnellSmith2011}.  The evidence from our ALMA and SMA observations that in G11.92$-$0.61 high-mass and low-mass stars are forming \emph{simultaneously}--e.g., are accreting at the same time, from the same clump-scale gas reservoir, as they approach their final masses--further supports dynamic cluster formation models in which high-mass stars form concurrently with their surrounding clusters.  

Our detection of accreting low-mass protostars in the outer reaches of the accretion reservoir, 0.2-0.3 pc from the central massive source MM1, emphasises the importance of sensitive, high-fidelity interferometric mosaics in addressing questions related to cluster formation, such as the relative birth order of low- and high-mass stars.  
Detailed comparison of our observational results with theoretical models will require synthetic observations of the numerical simulations, to enable direct comparisons, while  
investigating whether the onset of low- and high-mass star formation is coeval will require ALMA observations of a sample of protoclusters spanning a range of evolutionary stages.     

\section*{Acknowledgements}

CJC thanks M.~M. Dunham for helpful discussions on measuring outflow properties.  
This paper makes use of the following ALMA data: ADS/JAO.ALMA\#2013.1.00812.S.  ALMA is a
partnership of ESO (representing its member states), NSF (USA) and NINS (Japan), together with 
NRC (Canada), NSC and ASIAA (Taiwan), and KASI (Republic of Korea), in cooperation with the 
Republic of Chile. The Joint ALMA Observatory is operated by ESO, AUI/NRAO and NAOJ.  The National Radio Astronomy Observatory is a facility of the National Science Foundation operated under cooperative agreement by Associated Universities, Inc. This research has made use of NASA's Astrophysics Data System Bibliographic Services, Astropy, a community-developed core Python package for Astronomy \citep{astropy}, and APLpy, an open-source plotting package for Python hosted at http://aplpy.github.com.  The ATLASGAL project is a collaboration between the Max-Planck-Gesellschaft, the European Southern Observatory (ESO) and the Universidad de Chile. It includes projects E-181.C-0885, E-078.F-9040(A), M-079.C-9501(A), M-081.C-9501(A) plus Chilean data.
CJC acknowledges support from the STFC (grant number ST/M001296/1).  RJS gratefully acknowledges support from the RAS through their Norman Lockyer Fellowship.
JMDK gratefully acknowledges financial support in the form of an Emmy Noether Research Group from the Deutsche Forschungsgemeinschaft (DFG), grant number KR4801/1-1.
IAB acknowledges funding from the European Research
Council for the FP7 ERC advanced grant project ECOGAL.




\bibliographystyle{mnras}
\bibliography{g11_alma_mnras} 

\newcommand{\noop}[1]{}
\begin{thebibliography}{}
\makeatletter
\relax
\def\mn@urlcharsother{\let\do\@makeother \do\$\do\&\do\#\do\^\do\_\do\%\do\~}
\def\mn@doi{\begingroup\mn@urlcharsother \@ifnextchar [ {\mn@doi@}
  {\mn@doi@[]}}
\def\mn@doi@[#1]#2{\def\@tempa{#1}\ifx\@tempa\@empty \href
  {http://dx.doi.org/#2} {doi:#2}\else \href {http://dx.doi.org/#2} {#1}\fi
  \endgroup}
\def\mn@eprint#1#2{\mn@eprint@#1:#2::\@nil}
\def\mn@eprint@arXiv#1{\href {http://arxiv.org/abs/#1} {{\tt arXiv:#1}}}
\def\mn@eprint@dblp#1{\href {http://dblp.uni-trier.de/rec/bibtex/#1.xml}
  {dblp:#1}}
\def\mn@eprint@#1:#2:#3:#4\@nil{\def\@tempa {#1}\def\@tempb {#2}\def\@tempc
  {#3}\ifx \@tempc \@empty \let \@tempc \@tempb \let \@tempb \@tempa \fi \ifx
  \@tempb \@empty \def\@tempb {arXiv}\fi \@ifundefined
  {mn@eprint@\@tempb}{\@tempb:\@tempc}{\expandafter \expandafter \csname
  mn@eprint@\@tempb\endcsname \expandafter{\@tempc}}}

\bibitem[\protect\citeauthoryear{{Andrews}, {Wilner}, {Hughes}, {Qi}  \&
  {Dullemond}}{{Andrews} et~al.}{2010}]{Andrews2010}
{Andrews} S.~M.,  {Wilner} D.~J.,  {Hughes} A.~M.,  {Qi} C.,   {Dullemond}
  C.~P.,  2010, \mn@doi [\apj] {10.1088/0004-637X/723/2/1241}, \href
  {http://adsabs.harvard.edu/abs/2010ApJ...723.1241A} {723, 1241}

\bibitem[\protect\citeauthoryear{{Arce} \& {Sargent}}{{Arce} \&
  {Sargent}}{2006}]{Arce2006}
{Arce} H.~G.,  {Sargent} A.~I.,  2006, \mn@doi [\apj] {10.1086/505104}, \href
  {http://adsabs.harvard.edu/abs/2006ApJ...646.1070A} {646, 1070}

\bibitem[\protect\citeauthoryear{{Arce}, {Shepherd}, {Gueth}, {Lee},
  {Bachiller}, {Rosen}  \& {Beuther}}{{Arce} et~al.}{2007}]{Arce2007}
{Arce} H.~G.,  {Shepherd} D.,  {Gueth} F.,  {Lee} C.-F.,  {Bachiller} R.,
  {Rosen} A.,   {Beuther} H.,  2007, Protostars and Planets V, \href
  {http://adsabs.harvard.edu/abs/2007prpl.conf..245A} {pp 245--260}

\bibitem[\protect\citeauthoryear{{Astropy Collaboration} et~al.,}{{Astropy
  Collaboration} et~al.}{2013}]{astropy}
{Astropy Collaboration} et~al., 2013, \mn@doi [\aap]
  {10.1051/0004-6361/201322068}, \href
  {http://adsabs.harvard.edu/abs/2013A\&A...558A..33A} {558, A33}

\bibitem[\protect\citeauthoryear{{Bachiller} \& {P{\'e}rez
  Guti{\'e}rrez}}{{Bachiller} \& {P{\'e}rez
  Guti{\'e}rrez}}{1997}]{Bachiller1997}
{Bachiller} R.,  {P{\'e}rez Guti{\'e}rrez} M.,  1997, \mn@doi [\apjl]
  {10.1086/310877}, \href {http://adsabs.harvard.edu/abs/1997ApJ...487L..93B}
  {487, L93}

\bibitem[\protect\citeauthoryear{{Beuther}, {Schilke}, {Sridharan}, {Menten},
  {Walmsley}  \& {Wyrowski}}{{Beuther} et~al.}{2002a}]{Beuther2002}
{Beuther} H.,  {Schilke} P.,  {Sridharan} T.~K.,  {Menten} K.~M.,  {Walmsley}
  C.~M.,   {Wyrowski} F.,  2002a, \mn@doi [\aap] {10.1051/0004-6361:20011808},
  \href {http://adsabs.harvard.edu/abs/2002A%26A...383..892B} {383, 892}

\bibitem[\protect\citeauthoryear{{Beuther}, {Schilke}, {Menten}, {Motte},
  {Sridharan}  \& {Wyrowski}}{{Beuther} et~al.}{2002b}]{Beuther2002core}
{Beuther} H.,  {Schilke} P.,  {Menten} K.~M.,  {Motte} F.,  {Sridharan} T.~K.,
   {Wyrowski} F.,  2002b, \mn@doi [\apj] {10.1086/338334}, \href
  {http://adsabs.harvard.edu/abs/2002ApJ...566..945B} {566, 945}

\bibitem[\protect\citeauthoryear{{Bonnell} \& {Smith}}{{Bonnell} \&
  {Smith}}{2011}]{BonnellSmith2011}
{Bonnell} I.~A.,  {Smith} R.~J.,  2011, in {Alves} J.,  {Elmegreen} B.~G.,
  {Girart} J.~M.,   {Trimble} V.,  eds,  IAU Symposium Vol. 270, Computational
  Star Formation. pp 57--64, \mn@doi{10.1017/S1743921311000184}

\bibitem[\protect\citeauthoryear{{Bonnell}, {Vine}  \& {Bate}}{{Bonnell}
  et~al.}{2004}]{Bonnell2004}
{Bonnell} I.~A.,  {Vine} S.~G.,   {Bate} M.~R.,  2004, \mn@doi [\mnras]
  {10.1111/j.1365-2966.2004.07543.x}, \href
  {http://adsabs.harvard.edu/abs/2004MNRAS.349..735B} {349, 735}

\bibitem[\protect\citeauthoryear{{Bontemps}, {Andre}, {Terebey}  \&
  {Cabrit}}{{Bontemps} et~al.}{1996}]{Bontemps1996}
{Bontemps} S.,  {Andre} P.,  {Terebey} S.,   {Cabrit} S.,  1996, \aap, \href
  {http://adsabs.harvard.edu/abs/1996A\&A...311..858B} {311, 858}

\bibitem[\protect\citeauthoryear{{Breen} \& {Ellingsen}}{{Breen} \&
  {Ellingsen}}{2011}]{Breen2011}
{Breen} S.~L.,  {Ellingsen} S.~P.,  2011, \mn@doi [\mnras]
  {10.1111/j.1365-2966.2011.19020.x}, \href
  {http://adsabs.harvard.edu/abs/2011MNRAS.416..178B} {416, 178}

\bibitem[\protect\citeauthoryear{{Brogan}, {Hunter}, {Cyganowski}, {Friesen},
  {Chandler}  \& {Indebetouw}}{{Brogan} et~al.}{2011}]{Brogan2011}
{Brogan} C.~L.,  {Hunter} T.~R.,  {Cyganowski} C.~J.,  {Friesen} R.~K.,
  {Chandler} C.~J.,   {Indebetouw} R.,  2011, \mn@doi [\apjl]
  {10.1088/2041-8205/739/1/L16}, \href
  {http://adsabs.harvard.edu/abs/2011ApJ...739L..16B} {739, L16}

\bibitem[\protect\citeauthoryear{{Brogan}, {Hunter}, {Cyganowski}, {Chandler},
  {Friesen}  \& {Indebetouw}}{{Brogan} et~al.}{2016}]{Brogan2016}
{Brogan} C.~L.,  {Hunter} T.~R.,  {Cyganowski} C.~J.,  {Chandler} C.~J.,
  {Friesen} R.,   {Indebetouw} R.,  2016, \mn@doi [\apj]
  {10.3847/0004-637X/832/2/187}, \href
  {http://adsabs.harvard.edu/abs/2016ApJ...832..187B} {832, 187}

\bibitem[\protect\citeauthoryear{{Cabrit} \& {Bertout}}{{Cabrit} \&
  {Bertout}}{1992}]{Cabrit1992}
{Cabrit} S.,  {Bertout} C.,  1992, \aap, \href
  {http://adsabs.harvard.edu/abs/1992A\&A...261..274C} {261, 274}

\bibitem[\protect\citeauthoryear{{Contreras} et~al.,}{{Contreras}
  et~al.}{2013}]{Contreras2013}
{Contreras} Y.,  et~al., 2013, \mn@doi [\aap] {10.1051/0004-6361/201220155},
  \href {http://adsabs.harvard.edu/abs/2013A\&A...549A..45C} {549, A45}

\bibitem[\protect\citeauthoryear{{Csengeri} et~al.,}{{Csengeri}
  et~al.}{2014}]{Csengeri2014}
{Csengeri} T.,  et~al., 2014, \mn@doi [\aap] {10.1051/0004-6361/201322434},
  \href {http://adsabs.harvard.edu/abs/2014A\&A...565A..75C} {565, A75}

\bibitem[\protect\citeauthoryear{{Cunningham}, {Lumsden}, {Cyganowski}, {Maud}
  \& {Purcell}}{{Cunningham} et~al.}{2016}]{Cunningham2016}
{Cunningham} N.,  {Lumsden} S.~L.,  {Cyganowski} C.~J.,  {Maud} L.~T.,
  {Purcell} C.,  2016, \mn@doi [\mnras] {10.1093/mnras/stw359}, \href
  {http://adsabs.harvard.edu/abs/2016MNRAS.458.1742C} {458, 1742}

\bibitem[\protect\citeauthoryear{{Cyganowski}, {Brogan}  \&
  {Hunter}}{{Cyganowski} et~al.}{2007}]{Cyganowski2007}
{Cyganowski} C.~J.,  {Brogan} C.~L.,   {Hunter} T.~R.,  2007, \mn@doi [\aj]
  {10.1086/518740}, \href {http://adsabs.harvard.edu/abs/2007AJ....134..346C}
  {134, 346}

\bibitem[\protect\citeauthoryear{{Cyganowski} et~al.,}{{Cyganowski}
  et~al.}{2008}]{C08}
{Cyganowski} C.~J.,  et~al., 2008, \mn@doi [\aj]
  {10.1088/0004-6256/136/6/2391}, \href
  {http://adsabs.harvard.edu/abs/2008AJ....136.2391C} {136, 2391}

\bibitem[\protect\citeauthoryear{{Cyganowski}, {Brogan}, {Hunter}  \&
  {Churchwell}}{{Cyganowski} et~al.}{2009}]{C09}
{Cyganowski} C.~J.,  {Brogan} C.~L.,  {Hunter} T.~R.,   {Churchwell} E.,  2009,
  \mn@doi [\apj] {10.1088/0004-637X/702/2/1615}, \href
  {http://adsabs.harvard.edu/abs/2009ApJ...702.1615C} {702, 1615}

\bibitem[\protect\citeauthoryear{{Cyganowski}, {Brogan}, {Hunter}, {Churchwell}
   \& {Zhang}}{{Cyganowski} et~al.}{2011a}]{C11sma}
{Cyganowski} C.~J.,  {Brogan} C.~L.,  {Hunter} T.~R.,  {Churchwell} E.,
  {Zhang} Q.,  2011a, \mn@doi [\apj] {10.1088/0004-637X/729/2/124}, \href
  {http://adsabs.harvard.edu/abs/2011ApJ...729..124C} {729, 124}

\bibitem[\protect\citeauthoryear{{Cyganowski}, {Brogan}, {Hunter}  \&
  {Churchwell}}{{Cyganowski} et~al.}{2011b}]{C11vla}
{Cyganowski} C.~J.,  {Brogan} C.~L.,  {Hunter} T.~R.,   {Churchwell} E.,
  2011b, \mn@doi [\apj] {10.1088/0004-637X/743/1/56}, \href
  {http://adsabs.harvard.edu/abs/2011ApJ...743...56C} {743, 56}

\bibitem[\protect\citeauthoryear{{Cyganowski}, {Brogan}, {Hunter}, {Zhang},
  {Friesen}, {Indebetouw}  \& {Chandler}}{{Cyganowski}
  et~al.}{2012}]{Cyganowski2012}
{Cyganowski} C.~J.,  {Brogan} C.~L.,  {Hunter} T.~R.,  {Zhang} Q.,  {Friesen}
  R.~K.,  {Indebetouw} R.,   {Chandler} C.~J.,  2012, \mn@doi [\apjl]
  {10.1088/2041-8205/760/2/L20}, \href
  {http://adsabs.harvard.edu/abs/2012ApJ...760L..20C} {760, L20}

\bibitem[\protect\citeauthoryear{{Cyganowski} et~al.,}{{Cyganowski}
  et~al.}{2014}]{Cyganowski2014}
{Cyganowski} C.~J.,  et~al., 2014, \mn@doi [\apjl]
  {10.1088/2041-8205/796/1/L2}, \href
  {http://adsabs.harvard.edu/abs/2014ApJ...796L...2C} {796, L2}

\bibitem[\protect\citeauthoryear{{Downes} \& {Cabrit}}{{Downes} \&
  {Cabrit}}{2007}]{Downes2007}
{Downes} T.~P.,  {Cabrit} S.,  2007, \mn@doi [\aap]
  {10.1051/0004-6361:20066921}, \href
  {http://adsabs.harvard.edu/abs/2007A\&A...471..873D} {471, 873}

\bibitem[\protect\citeauthoryear{{Duarte-Cabral}, {Bontemps}, {Motte},
  {Hennemann}, {Schneider}  \& {Andr{\'e}}}{{Duarte-Cabral}
  et~al.}{2013}]{DuarteCabral2013}
{Duarte-Cabral} A.,  {Bontemps} S.,  {Motte} F.,  {Hennemann} M.,  {Schneider}
  N.,   {Andr{\'e}} P.,  2013, \mn@doi [\aap] {10.1051/0004-6361/201321393},
  \href {http://adsabs.harvard.edu/abs/2013A\&A...558A.125D} {558, A125}

\bibitem[\protect\citeauthoryear{{Dunham}, {Arce}, {Mardones}, {Lee},
  {Matthews}, {Stutz}  \& {Williams}}{{Dunham} et~al.}{2014}]{Dunham2014}
{Dunham} M.~M.,  {Arce} H.~G.,  {Mardones} D.,  {Lee} J.-E.,  {Matthews} B.~C.,
   {Stutz} A.~M.,   {Williams} J.~P.,  2014, \mn@doi [\apj]
  {10.1088/0004-637X/783/1/29}, \href
  {http://adsabs.harvard.edu/abs/2014ApJ...783...29D} {783, 29}

\bibitem[\protect\citeauthoryear{{Enoch}, {Evans}, {Sargent}, {Glenn},
  {Rosolowsky}  \& {Myers}}{{Enoch} et~al.}{2008}]{Enoch2008}
{Enoch} M.~L.,  {Evans} II N.~J.,  {Sargent} A.~I.,  {Glenn} J.,  {Rosolowsky}
  E.,   {Myers} P.,  2008, \mn@doi [\apj] {10.1086/589963}, \href
  {http://adsabs.harvard.edu/abs/2008ApJ...684.1240E} {684, 1240}

\bibitem[\protect\citeauthoryear{{Enoch} et~al.,}{{Enoch}
  et~al.}{2011}]{Enoch2011}
{Enoch} M.~L.,  et~al., 2011, \mn@doi [\apjs] {10.1088/0067-0049/195/2/21},
  \href {http://adsabs.harvard.edu/abs/2011ApJS..195...21E} {195, 21}

\bibitem[\protect\citeauthoryear{{Fa{\'u}ndez}, {Bronfman}, {Garay}, {Chini},
  {Nyman}  \& {May}}{{Fa{\'u}ndez} et~al.}{2004}]{Faundez2004}
{Fa{\'u}ndez} S.,  {Bronfman} L.,  {Garay} G.,  {Chini} R.,  {Nyman} L.-{\AA}.,
    {May} J.,  2004, \mn@doi [\aap] {10.1051/0004-6361:20035755}, \href
  {http://adsabs.harvard.edu/abs/2004A\&A...426...97F} {426, 97}

\bibitem[\protect\citeauthoryear{{Foster} et~al.,}{{Foster}
  et~al.}{2014}]{Foster2014}
{Foster} J.~B.,  et~al., 2014, \mn@doi [\apj] {10.1088/0004-637X/791/2/108},
  \href {http://adsabs.harvard.edu/abs/2014ApJ...791..108F} {791, 108}

\bibitem[\protect\citeauthoryear{{Frank} et~al.,}{{Frank}
  et~al.}{2014}]{Frank2014}
{Frank} A.,  et~al., 2014, \mn@doi [Protostars and Planets VI]
  {10.2458/azu_uapress_9780816531240-ch020}, \href
  {http://adsabs.harvard.edu/abs/2014prpl.conf..451F} {pp 451--474}

\bibitem[\protect\citeauthoryear{{Hatchell} \& {Dunham}}{{Hatchell} \&
  {Dunham}}{2009}]{Hatchell2009}
{Hatchell} J.,  {Dunham} M.~M.,  2009, \mn@doi [\aap]
  {10.1051/0004-6361/200911818}, \href
  {http://adsabs.harvard.edu/abs/2009A\&A...502..139H} {502, 139}

\bibitem[\protect\citeauthoryear{{Herbig}}{{Herbig}}{1962}]{Herbig1962}
{Herbig} G.~H.,  1962, Advances in Astronomy and Astrophysics, \href
  {http://adsabs.harvard.edu/abs/1962AdA\&A...1...47H} {1, 47}

\bibitem[\protect\citeauthoryear{{Hofner} \& {Churchwell}}{{Hofner} \&
  {Churchwell}}{1996}]{HC96}
{Hofner} P.,  {Churchwell} E.,  1996, \aaps, \href
  {http://cdsads.u-strasbg.fr/abs/1996A\&AS..120..283H} {120, 283}

\bibitem[\protect\citeauthoryear{{Hogerheijde}, {van Dishoeck}, {Salverda}  \&
  {Blake}}{{Hogerheijde} et~al.}{1999}]{Hogerheijde1999}
{Hogerheijde} M.~R.,  {van Dishoeck} E.~F.,  {Salverda} J.~M.,   {Blake} G.~A.,
   1999, \mn@doi [\apj] {10.1086/306844}, \href
  {http://adsabs.harvard.edu/abs/1999ApJ...513..350H} {513, 350}

\bibitem[\protect\citeauthoryear{{Hunter}, {Brogan}, {Indebetouw}  \&
  {Cyganowski}}{{Hunter} et~al.}{2008}]{Hunter2008}
{Hunter} T.~R.,  {Brogan} C.~L.,  {Indebetouw} R.,   {Cyganowski} C.~J.,  2008,
  \mn@doi [\apj] {10.1086/588016}, \href
  {http://adsabs.harvard.edu/abs/2008ApJ...680.1271H} {680, 1271}

\bibitem[\protect\citeauthoryear{{Hunter}, {Brogan}, {Cyganowski}  \&
  {Young}}{{Hunter} et~al.}{2014}]{Hunter2014}
{Hunter} T.~R.,  {Brogan} C.~L.,  {Cyganowski} C.~J.,   {Young} K.~H.,  2014,
  \mn@doi [\apj] {10.1088/0004-637X/788/2/187}, \href
  {http://adsabs.harvard.edu/abs/2014ApJ...788..187H} {788, 187}

\bibitem[\protect\citeauthoryear{{Hunter}, {Brogan}, {Cyganowski}  \&
  {Schnee}}{{Hunter} et~al.}{2016}]{Hunter2016}
{Hunter} T.~R.,  {Brogan} C.~L.,  {Cyganowski} C.~J.,   {Schnee} S.,  2016, in
  EAS Publications Series. pp 285--286, \mn@doi{10.1051/eas/1575057}

\bibitem[\protect\citeauthoryear{{Ilee}, {Cyganowski}, {Nazari}, {Hunter},
  {Brogan}, {Forgan}  \& {Zhang}}{{Ilee} et~al.}{2016}]{Ilee2016}
{Ilee} J.~D.,  {Cyganowski} C.~J.,  {Nazari} P.,  {Hunter} T.~R.,  {Brogan}
  C.~L.,  {Forgan} D.~H.,   {Zhang} Q.,  2016, \mn@doi [\mnras]
  {10.1093/mnras/stw1912}, \href
  {http://adsabs.harvard.edu/abs/2016MNRAS.462.4386I} {462, 4386}

\bibitem[\protect\citeauthoryear{{Kauffmann}, {Bertoldi}, {Bourke}, {Evans}  \&
  {Lee}}{{Kauffmann} et~al.}{2008}]{Kauffmann2008}
{Kauffmann} J.,  {Bertoldi} F.,  {Bourke} T.~L.,  {Evans} II N.~J.,   {Lee}
  C.~W.,  2008, \mn@doi [\aap] {10.1051/0004-6361:200809481}, \href
  {http://adsabs.harvard.edu/abs/2008A\&A...487..993K} {487, 993}

\bibitem[\protect\citeauthoryear{{Kirk}, {Johnstone}  \& {Di Francesco}}{{Kirk}
  et~al.}{2006}]{Kirk2006}
{Kirk} H.,  {Johnstone} D.,   {Di Francesco} J.,  2006, \mn@doi [\apj]
  {10.1086/503193}, \href {http://adsabs.harvard.edu/abs/2006ApJ...646.1009K}
  {646, 1009}

\bibitem[\protect\citeauthoryear{{Kruijssen}}{{Kruijssen}}{2012}]{Kruijssen201%
2}
{Kruijssen} J.~M.~D.,  2012, \mn@doi [\mnras]
  {10.1111/j.1365-2966.2012.21923.x}, \href
  {http://adsabs.harvard.edu/abs/2012MNRAS.426.3008K} {426, 3008}

\bibitem[\protect\citeauthoryear{{Kumar}, {Keto}  \& {Clerkin}}{{Kumar}
  et~al.}{2006}]{Kumar2006}
{Kumar} M.~S.~N.,  {Keto} E.,   {Clerkin} E.,  2006, \mn@doi [\aap]
  {10.1051/0004-6361:20053104}, \href
  {http://adsabs.harvard.edu/abs/2006A\&A...449.1033K} {449, 1033}

\bibitem[\protect\citeauthoryear{{Kurtz}, {Hofner}  \& {{\'A}lvarez}}{{Kurtz}
  et~al.}{2004}]{Kurtz2004}
{Kurtz} S.,  {Hofner} P.,   {{\'A}lvarez} C.~V.,  2004, \mn@doi [\apjs]
  {10.1086/423956}, \href {http://adsabs.harvard.edu/abs/2004ApJS..155..149K}
  {155, 149}

\bibitem[\protect\citeauthoryear{{Lada} \& {Lada}}{{Lada} \&
  {Lada}}{2003}]{Lada2003}
{Lada} C.~J.,  {Lada} E.~A.,  2003, \mn@doi [\araa]
  {10.1146/annurev.astro.41.011802.094844}, \href
  {http://adsabs.harvard.edu/abs/2003ARA\&A..41...57L} {41, 57}

\bibitem[\protect\citeauthoryear{{Lee}, {Takami}, {Duan}, {Karr}, {Su}, {Liu},
  {Froebrich}  \& {Yeh}}{{Lee} et~al.}{2012}]{Lee2012}
{Lee} H.-T.,  {Takami} M.,  {Duan} H.-Y.,  {Karr} J.,  {Su} Y.-N.,  {Liu}
  S.-Y.,  {Froebrich} D.,   {Yeh} C.~C.,  2012, \mn@doi [\apjs]
  {10.1088/0067-0049/200/1/2}, \href
  {http://adsabs.harvard.edu/abs/2012ApJS..200....2L} {200, 2}

\bibitem[\protect\citeauthoryear{{Lee} et~al.,}{{Lee} et~al.}{2013}]{Lee2013}
{Lee} H.-T.,  et~al., 2013, \mn@doi [\apjs] {10.1088/0067-0049/208/2/23}, \href
  {http://adsabs.harvard.edu/abs/2013ApJS..208...23L} {208, 23}

\bibitem[\protect\citeauthoryear{{Mangum} \& {Shirley}}{{Mangum} \&
  {Shirley}}{2015}]{MangumShirley2015}
{Mangum} J.~G.,  {Shirley} Y.~L.,  2015, \mn@doi [\pasp] {10.1086/680323},
  \href {http://adsabs.harvard.edu/abs/2015PASP..127..266M} {127, 266}

\bibitem[\protect\citeauthoryear{{Maud}, {Lumsden}, {Moore}, {Mottram},
  {Urquhart}  \& {Cicchini}}{{Maud} et~al.}{2015a}]{Maud2015core}
{Maud} L.~T.,  {Lumsden} S.~L.,  {Moore} T.~J.~T.,  {Mottram} J.~C.,
  {Urquhart} J.~S.,   {Cicchini} A.,  2015a, \mn@doi [\mnras]
  {10.1093/mnras/stv1334}, \href
  {http://adsabs.harvard.edu/abs/2015MNRAS.452..637M} {452, 637}

\bibitem[\protect\citeauthoryear{{Maud}, {Moore}, {Lumsden}, {Mottram},
  {Urquhart}  \& {Hoare}}{{Maud} et~al.}{2015b}]{Maud2015out}
{Maud} L.~T.,  {Moore} T.~J.~T.,  {Lumsden} S.~L.,  {Mottram} J.~C.,
  {Urquhart} J.~S.,   {Hoare} M.~G.,  2015b, \mn@doi [\mnras]
  {10.1093/mnras/stv1635}, \href
  {http://adsabs.harvard.edu/abs/2015MNRAS.453..645M} {453, 645}

\bibitem[\protect\citeauthoryear{{Maury}, {Andr{\'e}}  \& {Li}}{{Maury}
  et~al.}{2009}]{Maury2009}
{Maury} A.~J.,  {Andr{\'e}} P.,   {Li} Z.-Y.,  2009, \mn@doi [\aap]
  {10.1051/0004-6361/200811442}, \href
  {http://adsabs.harvard.edu/abs/2009A\&A...499..175M} {499, 175}

\bibitem[\protect\citeauthoryear{{McKee} \& {Tan}}{{McKee} \&
  {Tan}}{2002}]{MT2002}
{McKee} C.~F.,  {Tan} J.~C.,  2002, \nat, \href
  {http://adsabs.harvard.edu/abs/2002Natur.416...59M} {416, 59}

\bibitem[\protect\citeauthoryear{{McKee} \& {Tan}}{{McKee} \&
  {Tan}}{2003}]{MT2003}
{McKee} C.~F.,  {Tan} J.~C.,  2003, \mn@doi [\apj] {10.1086/346149}, \href
  {http://adsabs.harvard.edu/abs/2003ApJ...585..850M} {585, 850}

\bibitem[\protect\citeauthoryear{{Minier}, {Ellingsen}, {Norris}  \&
  {Booth}}{{Minier} et~al.}{2003}]{Minier2003}
{Minier} V.,  {Ellingsen} S.~P.,  {Norris} R.~P.,   {Booth} R.~S.,  2003,
  \mn@doi [\aap] {10.1051/0004-6361:20030465}, \href
  {http://adsabs.harvard.edu/abs/2003A\&A...403.1095M} {403, 1095}

\bibitem[\protect\citeauthoryear{{Moscadelli} et~al.,}{{Moscadelli}
  et~al.}{2016}]{Moscadelli2016}
{Moscadelli} L.,  et~al., 2016, \mn@doi [\aap] {10.1051/0004-6361/201526238},
  \href {http://adsabs.harvard.edu/abs/2016A\&A...585A..71M} {585, A71}

\bibitem[\protect\citeauthoryear{{M{\"u}ller}, {Thorwirth}, {Roth}  \&
  {Winnewisser}}{{M{\"u}ller} et~al.}{2001}]{Muller2001}
{M{\"u}ller} H.~S.~P.,  {Thorwirth} S.,  {Roth} D.~A.,   {Winnewisser} G.,
  2001, \mn@doi [\aap] {10.1051/0004-6361:20010367}, \href
  {http://adsabs.harvard.edu/abs/2001A\&A...370L..49M} {370, L49}

\bibitem[\protect\citeauthoryear{{M{\"u}ller}, {Schl{\"o}der}, {Stutzki}  \&
  {Winnewisser}}{{M{\"u}ller} et~al.}{2005}]{Muller2005}
{M{\"u}ller} H.~S.~P.,  {Schl{\"o}der} F.,  {Stutzki} J.,   {Winnewisser} G.,
  2005, \mn@doi [Journal of Molecular Structure]
  {10.1016/j.molstruc.2005.01.027}, \href
  {http://adsabs.harvard.edu/abs/2005JMoSt.742..215M} {742, 215}

\bibitem[\protect\citeauthoryear{{Myers}, {McKee}, {Cunningham}, {Klein}  \&
  {Krumholz}}{{Myers} et~al.}{2013}]{Myers2013}
{Myers} A.~T.,  {McKee} C.~F.,  {Cunningham} A.~J.,  {Klein} R.~I.,
  {Krumholz} M.~R.,  2013, \mn@doi [\apj] {10.1088/0004-637X/766/2/97}, \href
  {http://adsabs.harvard.edu/abs/2013ApJ...766...97M} {766, 97}

\bibitem[\protect\citeauthoryear{{Offner}, {Lee}, {Goodman}  \&
  {Arce}}{{Offner} et~al.}{2011}]{Offner2011}
{Offner} S.~S.~R.,  {Lee} E.~J.,  {Goodman} A.~A.,   {Arce} H.,  2011, \mn@doi
  [\apj] {10.1088/0004-637X/743/1/91}, \href
  {http://adsabs.harvard.edu/abs/2011ApJ...743...91O} {743, 91}

\bibitem[\protect\citeauthoryear{{Ossenkopf} \& {Henning}}{{Ossenkopf} \&
  {Henning}}{1994}]{OH94}
{Ossenkopf} V.,  {Henning} T.,  1994, \aap, \href
  {http://adsabs.harvard.edu/abs/1994A\&A...291..943O} {291, 943}

\bibitem[\protect\citeauthoryear{{Peters}, {Banerjee}, {Klessen}, {Mac Low},
  {Galv{\'a}n-Madrid}  \& {Keto}}{{Peters} et~al.}{2010a}]{Peters2010a}
{Peters} T.,  {Banerjee} R.,  {Klessen} R.~S.,  {Mac Low} M.-M.,
  {Galv{\'a}n-Madrid} R.,   {Keto} E.~R.,  2010a, \mn@doi [\apj]
  {10.1088/0004-637X/711/2/1017}, \href
  {http://adsabs.harvard.edu/abs/2010ApJ...711.1017P} {711, 1017}

\bibitem[\protect\citeauthoryear{{Peters}, {Klessen}, {Mac Low}  \&
  {Banerjee}}{{Peters} et~al.}{2010b}]{Peters2010b}
{Peters} T.,  {Klessen} R.~S.,  {Mac Low} M.-M.,   {Banerjee} R.,  2010b,
  \mn@doi [\apj] {10.1088/0004-637X/725/1/134}, \href
  {http://adsabs.harvard.edu/abs/2010ApJ...725..134P} {725, 134}

\bibitem[\protect\citeauthoryear{{Peters}, {Banerjee}, {Klessen}  \& {Mac
  Low}}{{Peters} et~al.}{2011}]{Peters2011}
{Peters} T.,  {Banerjee} R.,  {Klessen} R.~S.,   {Mac Low} M.-M.,  2011,
  \mn@doi [\apj] {10.1088/0004-637X/729/1/72}, \href
  {http://adsabs.harvard.edu/abs/2011ApJ...729...72P} {729, 72}

\bibitem[\protect\citeauthoryear{{Plambeck} \& {Menten}}{{Plambeck} \&
  {Menten}}{1990}]{Plambeck1990}
{Plambeck} R.~L.,  {Menten} K.~M.,  1990, \mn@doi [\apj] {10.1086/169437},
  \href {http://adsabs.harvard.edu/abs/1990ApJ...364..555P} {364, 555}

\bibitem[\protect\citeauthoryear{{Povich} \& {Whitney}}{{Povich} \&
  {Whitney}}{2010}]{Povich2010}
{Povich} M.~S.,  {Whitney} B.~A.,  2010, \mn@doi [\apjl]
  {10.1088/2041-8205/714/2/L285}, \href
  {http://adsabs.harvard.edu/abs/2010ApJ...714L.285P} {714, L285}

\bibitem[\protect\citeauthoryear{{Qiu}, {Zhang}, {Wu}  \& {Chen}}{{Qiu}
  et~al.}{2009}]{Qiu2009}
{Qiu} K.,  {Zhang} Q.,  {Wu} J.,   {Chen} H.-R.,  2009, \mn@doi [\apj]
  {10.1088/0004-637X/696/1/66}, \href
  {http://adsabs.harvard.edu/abs/2009ApJ...696...66Q} {696, 66}

\bibitem[\protect\citeauthoryear{{Rathborne}, {Lada}, {Muench}, {Alves},
  {Kainulainen}  \& {Lombardi}}{{Rathborne} et~al.}{2009}]{Rathborne2009}
{Rathborne} J.~M.,  {Lada} C.~J.,  {Muench} A.~A.,  {Alves} J.~F.,
  {Kainulainen} J.,   {Lombardi} M.,  2009, \mn@doi [\apj]
  {10.1088/0004-637X/699/1/742}, \href
  {http://adsabs.harvard.edu/abs/2009ApJ...699..742R} {699, 742}

\bibitem[\protect\citeauthoryear{{Remijan}}{{Remijan}}{2010}]{Remijan2010}
{Remijan} A.~J.,  2010, in American Astronomical Society Meeting Abstracts
  \#215. p.~568

\bibitem[\protect\citeauthoryear{{Richer}, {Shepherd}, {Cabrit}, {Bachiller}
  \& {Churchwell}}{{Richer} et~al.}{2000}]{Richer2000}
{Richer} J.~S.,  {Shepherd} D.~S.,  {Cabrit} S.,  {Bachiller} R.,
  {Churchwell} E.,  2000, Protostars and Planets IV, \href
  {http://adsabs.harvard.edu/abs/2000prpl.conf..867R} {p.~867}

\bibitem[\protect\citeauthoryear{{Rivilla}, {Jim{\'e}nez-Serra},
  {Mart{\'{\i}}n-Pintado}  \& {Sanz-Forcada}}{{Rivilla}
  et~al.}{2014}]{Rivilla2014}
{Rivilla} V.~M.,  {Jim{\'e}nez-Serra} I.,  {Mart{\'{\i}}n-Pintado} J.,
  {Sanz-Forcada} J.,  2014, \mn@doi [\mnras] {10.1093/mnras/stt1989}, \href
  {http://adsabs.harvard.edu/abs/2014MNRAS.437.1561R} {437, 1561}

\bibitem[\protect\citeauthoryear{{Sato} et~al.,}{{Sato}
  et~al.}{2014}]{Sato2014}
{Sato} M.,  et~al., 2014, \mn@doi [\apj] {10.1088/0004-637X/793/2/72}, \href
  {http://adsabs.harvard.edu/abs/2014ApJ...793...72S} {793, 72}

\bibitem[\protect\citeauthoryear{{Schuller} et~al.,}{{Schuller}
  et~al.}{2009}]{Schuller2009}
{Schuller} F.,  et~al., 2009, \mn@doi [\aap] {10.1051/0004-6361/200811568},
  \href {http://adsabs.harvard.edu/abs/2009A\&A...504..415S} {504, 415}

\bibitem[\protect\citeauthoryear{{Shepherd} \& {Churchwell}}{{Shepherd} \&
  {Churchwell}}{1996}]{Shepherd1996}
{Shepherd} D.~S.,  {Churchwell} E.,  1996, \mn@doi [\apj] {10.1086/178057},
  \href {http://adsabs.harvard.edu/abs/1996ApJ...472..225S} {472, 225}

\bibitem[\protect\citeauthoryear{{Simpson}, {Nutter}  \&
  {Ward-Thompson}}{{Simpson} et~al.}{2008}]{Simpson2008}
{Simpson} R.~J.,  {Nutter} D.,   {Ward-Thompson} D.,  2008, \mn@doi [\mnras]
  {10.1111/j.1365-2966.2008.13750.x}, \href
  {http://adsabs.harvard.edu/abs/2008MNRAS.391..205S} {391, 205}

\bibitem[\protect\citeauthoryear{{Smith}, {Longmore}  \& {Bonnell}}{{Smith}
  et~al.}{2009}]{Smith2009}
{Smith} R.~J.,  {Longmore} S.,   {Bonnell} I.,  2009, \mn@doi [\mnras]
  {10.1111/j.1365-2966.2009.15621.x}, \href
  {http://adsabs.harvard.edu/abs/2009MNRAS.400.1775S} {400, 1775}

\bibitem[\protect\citeauthoryear{{Smith}, {Glover}, {Klessen}  \&
  {Fuller}}{{Smith} et~al.}{2016}]{Smith2016}
{Smith} R.~J.,  {Glover} S.~C.~O.,  {Klessen} R.~S.,   {Fuller} G.~A.,  2016,
  \mn@doi [\mnras] {10.1093/mnras/stv2559}, \href
  {http://adsabs.harvard.edu/abs/2016MNRAS.455.3640S} {455, 3640}

\bibitem[\protect\citeauthoryear{{Tafalla}, {Santiago-Garc{\'{\i}}a}, {Hacar}
  \& {Bachiller}}{{Tafalla} et~al.}{2010}]{Tafalla2010}
{Tafalla} M.,  {Santiago-Garc{\'{\i}}a} J.,  {Hacar} A.,   {Bachiller} R.,
  2010, \mn@doi [\aap] {10.1051/0004-6361/201015158}, \href
  {http://adsabs.harvard.edu/abs/2010A\&A...522A..91T} {522, A91}

\bibitem[\protect\citeauthoryear{{Tan}, {Beltr{\'a}n}, {Caselli}, {Fontani},
  {Fuente}, {Krumholz}, {McKee}  \& {Stolte}}{{Tan} et~al.}{2014}]{Tan2014}
{Tan} J.~C.,  {Beltr{\'a}n} M.~T.,  {Caselli} P.,  {Fontani} F.,  {Fuente} A.,
  {Krumholz} M.~R.,  {McKee} C.~F.,   {Stolte} A.,  2014, \mn@doi [Protostars
  and Planets VI] {10.2458/azu_uapress_9780816531240-ch007}, \href
  {http://adsabs.harvard.edu/abs/2014prpl.conf..149T} {pp 149--172}

\bibitem[\protect\citeauthoryear{{Thompson}, {Hatchell}, {Walsh}, {MacDonald}
  \& {Millar}}{{Thompson} et~al.}{2006}]{Thompson2006}
{Thompson} M.~A.,  {Hatchell} J.,  {Walsh} A.~J.,  {MacDonald} G.~H.,
  {Millar} T.~J.,  2006, \mn@doi [\aap] {10.1051/0004-6361:20054383}, \href
  {http://adsabs.harvard.edu/abs/2006A\&A...453.1003T} {453, 1003}

\bibitem[\protect\citeauthoryear{{Voronkov}, {Caswell}, {Ellingsen}, {Breen},
  {Britton}, {Green}, {Sobolev}  \& {Walsh}}{{Voronkov}
  et~al.}{2012}]{Voronkov2012}
{Voronkov} M.~A.,  {Caswell} J.~L.,  {Ellingsen} S.~P.,  {Breen} S.~L.,
  {Britton} T.~R.,  {Green} J.~A.,  {Sobolev} A.~M.,   {Walsh} A.~J.,  2012, in
  {Booth} R.~S.,  {Vlemmings} W.~H.~T.,   {Humphreys} E.~M.~L.,  eds,  IAU
  Symposium Vol. 287, Cosmic Masers - from OH to H0. pp 433--440 (\mn@eprint
  {arXiv} {1203.5492}), \mn@doi{10.1017/S174392131200748X}

\bibitem[\protect\citeauthoryear{{Walsh}, {Macdonald}, {Alvey}, {Burton}  \&
  {Lee}}{{Walsh} et~al.}{2003}]{Walsh2003}
{Walsh} A.~J.,  {Macdonald} G.~H.,  {Alvey} N.~D.~S.,  {Burton} M.~G.,   {Lee}
  J.-K.,  2003, \mn@doi [\aap] {10.1051/0004-6361:20031191}, \href
  {http://adsabs.harvard.edu/abs/2003A\&A...410..597W} {410, 597}

\bibitem[\protect\citeauthoryear{{Wang}, {Li}, {Abel}  \& {Nakamura}}{{Wang}
  et~al.}{2010}]{Wang2010}
{Wang} P.,  {Li} Z.-Y.,  {Abel} T.,   {Nakamura} F.,  2010, \mn@doi [\apj]
  {10.1088/0004-637X/709/1/27}, \href
  {http://adsabs.harvard.edu/abs/2010ApJ...709...27W} {709, 27}

\bibitem[\protect\citeauthoryear{{Wang} et~al.,}{{Wang}
  et~al.}{2014}]{Wang2014}
{Wang} K.,  et~al., 2014, \mn@doi [\mnras] {10.1093/mnras/stu127}, \href
  {http://adsabs.harvard.edu/abs/2014MNRAS.439.3275W} {439, 3275}

\bibitem[\protect\citeauthoryear{{Wilner} \& {Welch}}{{Wilner} \&
  {Welch}}{1994}]{Wilner1994}
{Wilner} D.~J.,  {Welch} W.~J.,  1994, \mn@doi [\apj] {10.1086/174195}, \href
  {http://adsabs.harvard.edu/abs/1994ApJ...427..898W} {427, 898}

\bibitem[\protect\citeauthoryear{{Wu}, {Wei}, {Zhao}, {Shi}, {Yu}, {Qin}  \&
  {Huang}}{{Wu} et~al.}{2004}]{Wu2004}
{Wu} Y.,  {Wei} Y.,  {Zhao} M.,  {Shi} Y.,  {Yu} W.,  {Qin} S.,   {Huang} M.,
  2004, \mn@doi [\aap] {10.1051/0004-6361:20035767}, \href
  {http://adsabs.harvard.edu/abs/2004A\&A...426..503W} {426, 503}

\bibitem[\protect\citeauthoryear{{Yanagida} et~al.,}{{Yanagida}
  et~al.}{2014}]{ALMA_278_maser}
{Yanagida} T.,  et~al., 2014, \mn@doi [\apjl] {10.1088/2041-8205/794/1/L10},
  \href {http://adsabs.harvard.edu/abs/2014ApJ...794L..10Y} {794, L10}

\bibitem[\protect\citeauthoryear{{Zhang}, {Hunter}, {Brand}, {Sridharan},
  {Molinari}, {Kramer}  \& {Cesaroni}}{{Zhang} et~al.}{2001}]{Zhang2001}
{Zhang} Q.,  {Hunter} T.~R.,  {Brand} J.,  {Sridharan} T.~K.,  {Molinari} S.,
  {Kramer} M.~A.,   {Cesaroni} R.,  2001, \mn@doi [\apjl] {10.1086/320345},
  \href {http://adsabs.harvard.edu/abs/2001ApJ...552L.167Z} {552, L167}

\bibitem[\protect\citeauthoryear{{Zhang}, {Wang}, {Lu}  \&
  {Jim{\'e}nez-Serra}}{{Zhang} et~al.}{2015}]{Zhang2015}
{Zhang} Q.,  {Wang} K.,  {Lu} X.,   {Jim{\'e}nez-Serra} I.,  2015, \mn@doi
  [\apj] {10.1088/0004-637X/804/2/141}, \href
  {http://adsabs.harvard.edu/abs/2015ApJ...804..141Z} {804, 141}

\bibitem[\protect\citeauthoryear{{Zinchenko}, {Liu}, {Su}, {Kurtz}, {Ojha},
  {Samal}  \& {Ghosh}}{{Zinchenko} et~al.}{2012}]{Zinchenko2012}
{Zinchenko} I.,  {Liu} S.-Y.,  {Su} Y.-N.,  {Kurtz} S.,  {Ojha} D.~K.,  {Samal}
  M.~R.,   {Ghosh} S.~K.,  2012, \mn@doi [\apj] {10.1088/0004-637X/755/2/177},
  \href {http://adsabs.harvard.edu/abs/2012ApJ...755..177Z} {755, 177}

\bibitem[\protect\citeauthoryear{{de Villiers} et~al.,}{{de Villiers}
  et~al.}{2014}]{deVilliers2014}
{de Villiers} H.~M.,  et~al., 2014, \mn@doi [\mnras] {10.1093/mnras/stu1474},
  \href {http://adsabs.harvard.edu/abs/2014MNRAS.444..566D} {444, 566}

\bibitem[\protect\citeauthoryear{{van Dishoeck} \& {Blake}}{{van Dishoeck} \&
  {Blake}}{1998}]{vDB98}
{van Dishoeck} E.~F.,  {Blake} G.~A.,  1998, \mn@doi [\araa]
  {10.1146/annurev.astro.36.1.317}, \href
  {http://adsabs.harvard.edu/abs/1998ARA\&A..36..317V} {36, 317}

\bibitem[\protect\citeauthoryear{{van der Marel}, {Kristensen}, {Visser},
  {Mottram}, {Y{\i}ld{\i}z}  \& {van Dishoeck}}{{van der Marel}
  et~al.}{2013}]{vanderMarel2013}
{van der Marel} N.,  {Kristensen} L.~E.,  {Visser} R.,  {Mottram} J.~C.,
  {Y{\i}ld{\i}z} U.~A.,   {van Dishoeck} E.~F.,  2013, \mn@doi [\aap]
  {10.1051/0004-6361/201220717}, \href
  {http://adsabs.harvard.edu/abs/2013A\&A...556A..76V} {556, A76}

\makeatother
\end{thebibliography}



\bsp	
\label{lastpage}
\end{document}